\documentclass[usenatbib,useAMS,twocolumn]{mn2e}\usepackage[]{graphicx}\usepackage[]{color}
\makeatletter
\def\maxwidth{ %
  \ifdim\Gin@nat@width>\linewidth
    \linewidth
  \else
    \Gin@nat@width
  \fi
}
\makeatother

\definecolor{fgcolor}{rgb}{0.345, 0.345, 0.345}

\usepackage{framed}
\makeatletter
 {\par\unskip\endMakeFramed%
 \at@end@of@kframe}
\makeatother

\definecolor{shadecolor}{rgb}{.97, .97, .97}
\definecolor{messagecolor}{rgb}{0, 0, 0}
\definecolor{warningcolor}{rgb}{1, 0, 1}
\definecolor{errorcolor}{rgb}{1, 0, 0}
\newenvironment{knitrout}{}{} % an empty environment to be redefined in TeX

\usepackage{alltt}

\usepackage{aas_macros}
\usepackage{amsmath}
\usepackage{amssymb}
\usepackage{graphicx}
\usepackage{longtable}
\usepackage{mathptmx}
\usepackage{multirow}
\usepackage{url}
\usepackage[section]{placeins}

\pagerange{\pageref{firstpage}--\pageref{lastpage}} \pubyear{2014}

\IfFileExists{upquote.sty}{\usepackage{upquote}}{}
\begin{document}

\title[LT RINGO3 polarimeter calibration]{Calibration of
the Liverpool Telescope RINGO3 polarimeter}

\author[A. S\l{}owikowska et al.]{Aga S\l{}owikowska$^1$, 
%\footnotemark[1],%\thanks{e-mail:aga@astro.ia.uz.zgora.pl},
Krzysztof~Krzeszowski$^1$, %\footnotemark[1], %\thanks{e-mail:k.chriss@astro.ia.uz.zgora.pl},
Micha\l{}~\.Zejmo$^1$, %\footnotemark[1], %\thanks{e-mail:mzejmo@astro.ia.uz.zgora.pl},
Pablo Reig$^{2,3}$ 
\newauthor
and Iain Steele$^{4}$\\
$^1$ Janusz Gil Institute of Astronomy, University of Zielona G\'ora, Lubuska 2,
65-265 Zielona G\'ora, Poland\\
$^2$ Foundation for Research and Technology, 71110 Heraklion, Crete, Greece\\
$^3$ University of Crete, Physics Department, PO Box 2208, 710 03 Heraklion, Crete, Greece\\
$^4$ Astrophysics Research Institute, Liverpool John Moores University, L3 5RF, UK\\
}

\date{Released 2014}
\maketitle

\begin{abstract}
We present an analysis of polarimetric observations of standard stars performed over the period of more than three years with the RINGO3 polarimeter
mounted on the Liverpool Telescope. The main objective was to determine the instrumental polarisation
of the RINGO3 polarimeter in three spectral energy ranges:
blue (350--640~nm), green (650--760~nm) and red (770--1000~nm).
The observations were conducted between 2012 and 2016. The total time span
of 1126 days was split into five epochs due to the hardware changes to the observing system. 
Our results should be applied to calibrate all polarimetric observations performed with the RINGO3 polarimeter.
\end{abstract}

\label{firstpage}


\section{Introduction}
\label{sec:introduction}

Polarimetry is an essential complement to photometric and
spectroscopic studies because it provides additional constraints on existing 
theoretical models describing the physical processes standing behind
emission of radiation. Radiation from astrophysical sources as a rule shows some 
degree of polarisation. However, the polarised radiation is often only a small
fraction of the total radiation, with typical value of polarisation degree of a
few percent. Polarised radiation carries a wealth of information on the physical 
state and geometry of the emitting source and interacting interstellar
medium. Therefore, polarimetry often yields information that other methods
of observations can not provide. Diagnostics of various astrophysical emission
and radiation transfer phenomena are often based on polarimetric measurements.
Good examples include optical polarimetric measurements simultaneous with 
gamma-ray flares in the blazar 3C279 \citep{Abdo2010},
polarimetery of optical afterglow of gamma ray bursts \citep[e.g. GRB 120308A, ][]{Mundell2013},
dust grains in the debris disk of AU Microscopii \citep{Graham2007}, polarised light from 
atmospheres of the solar planets and in extreme cases of exoplanets
\citep[e.g.][]{Hansen1974}.  High time resolution polarimetry is also important,
for example in the case of fast rotating neutron stars such as the Crab pulsar \citep{Slowikowska2009}.  An example
of highly spatial resolution polarimetric observations of nebulae including 
highly polarised rotation powered pulsar nebulae such as the Crab nebula is
presented by \citet{Moran2013}.

A polarimeter measures the state of polarisation, or
some aspects of the state of polarisation, of a beam of radiation. Ideally, the
values of all four Stokes parameters should be determinable, together with
their variations with time, space, and wavelength. In practise, this is rarely
possible, at least for astronomical sources.  Most of the time only the degree
of linear polarisation and its direction are found.
There is a variety of the polarimetry measurement techniques. They range 
from the simplest one, i.e. looking through a polariser or the equivalent
at other wavelengths, to specialised use of time varying wave plates and 
detectors like CCDs or radio interferometry arrays. The precise form
of the instrument depends on the wavelength range for which it is designed.

Proper polarimeter calibration is the most critical issue for all polarimetric measurements
in order to get reliable results. It is achieved with long term monitoring
of polarised and zero--polarised standard stars. Observations of both types of standard stars are required to
determine instrumental polarisation and depolarisation. 
Calibration is a necessary step to be able to compare measurements taken with different instruments or the same instrument at different epochs.

In this paper we present our data analysis of a number of polarised and zero-polarised standard stars 
obtained with the RINGO3 polarimeter mounted on the Liverpool Telescope - the biggest optical, 
fully robotic telescope on the world,  as a function of time with the aim of providing useful 
information for other users of the instrument. We describe the three colour
RINGO3\footnote{\url{http://telescope.livjm.ac.uk/TelInst/Inst/RINGO3/}}
polarimeter in Sec.~\ref{sec:ringo}. The observations are described 
in Sec.~\ref{sec:obs} while the data analysis is described in Sec.~\ref{sec:data}. 
We present our results in Sec.~\ref{sec:results} and summarise our work in Sec.~\ref{sec:summary},
whereas in the Appendix we present some of the details 
of our calculations, figures showing the normalized Stokes parameters, PD and PA
as a function of time for BD~+59~389, BD~+64~106, G~191~B2B and HD~14069. Additionally, 
we also give there the tables with the coefficients of linear fits to the normalized 
Stokes q an u parameters for two zero-polarized standard stars.

\section{RINGO3 polarimeter}
\label{sec:ringo}

RINGO3 \citep{Arnold2012} is a fast-readout optical imaging polarimeter
at the Liverpool Telescope \citep{Steele2004}.   
Unlike the original RINGO\footnote{\url{http://telescope.livjm.ac.uk/TelInst/Inst/RINGO/}} 
which used deviating optics to spread the time varying polarised signal into rings, RINGO3 uses a fast readout camera to capture this signal as it changes in time. 
It is fed by a 45 degree
folding mirror from the telescope main beam.  The first optical element in the
system is a field lens that places the telescope pupil close
to the position of the dichroic mirrors farther down the optical path.  This serves
to reduce the vignetting in the instrument.  Following this is a high quality wire grid polariser that
rotates approximately once per second.  It is this time varying signal that the instrument
records in order to measure the polarisation of light entering the instrument.
Following the rotating polariser, a collimator lens is used to collimate the beam. 
A pair of dichroic mirrors then splits the beam into three for simultaneous polarised
imaging in three wavebands with separate camera lens and detector systems with approximate
wavelength ranges of: blue 350--640 nm, green 650--760 nm and red 770--1000 nm.
The colours of the RINGO3 cameras therefore approximately correspond to the B+V, R and I Johnson 
filters respectively.
Each camera receives eight exposures per polariser rotation.  These exposures
are electronically synchronised with the phase of the polariser's rotation. Therefore
in a typical one minute exposure the instrument will produce $\sim60$ frames at each
phase (i.e. a total of $\sim 480$ frames).  All frames for each matched phase are then stacked to
obtain a single image at each phase of the polariser's rotation for the purposes of data analysis. 

The detectors comprise a $512\times512$ pixel electron multiplying CCD 
(EMCCD) cameras with negligible dark current.  
These yield a field of view of $\sim4-5$ arcmin diameter.  The field
of view and pixel scale varies slightly for each camera due to their slightly different
optical arrangements.  Tests on the highly polarised twilight sky 
show that there is no significant dependence of measured polarisation with position in the
unvignetted field.

The EMCCD gain can be set to values of 5, 20 or 100.  For sources fainter than $V\sim10$ a
gain of 100 is used.  For brighter sources a gain of 20 is generally employed. An EMGAIN of 100
corresponds to a gain value of around 0.32 electrons/ADU in a single $\sim125$-ms
frame. As described above the data pipeline automatically stacks images to create a mean frame
from a given polaroid phase. Therefore, the final gain of an image can be calculated
by multiplying 0.32 by the number of frames that were stacked.

For proper data calibration two types of polarimetric standard stars need to be observed.
By default polarised as well as zero-polarised standard stars are observed robotically each night.
The zero-polarised standard stars allow correction of the instrumental polarisation,
while the polarised standard stars correct for the instrumental depolarisation.
The Liverpool Telescope has an altitude-azimuth mount with associated Cassegrain rotator.
To minimise a potential source of systematic error these standard star observations are 
obtained at the same rotator mount position angle as the science data. In general the mount position
angle of zero is used for all RINGO3 observations (science and standard stars). To calculate the
true sky position angle it is necessary to take into account the sky position angle
and the mechanical mount position. Both values are stored in the FITS files headers
as \textit{ROTSKYPA} and \textit{ROTANGLE}, respectively.

\section{Observations}
\label{sec:obs}

Both zero and polarised standard stars have been and are regularly scheduled and observed during most nights with RINGO3. 
In this paper we analyse data from 2012, December 7th (56268 MJD) until 2016, January 7th (57394 MJD), i.e. throughout whole RINGO3 life cycle after 
the commissioning. The data were obtained from the public archive available on the LT web page\footnote{\url{http://telescope.livjm.ac.uk/cgi-bin/lt_search}}.
Within that time span there were four hardware changes, one of which also coincided with the re-aluminisation
of the telescope's primary and secondary mirrors, i.e.:

\begin{description}

\item[\textbf{2013-01-23} (56315 MJD)] On this date the field lens was changed to optimise the vignetting.  In the process of doing this the polariser was rotated relative to the electronic synchronisation reference sensor. 

\item[\textbf{2013-12-12} (56638 MJD)] On this date a Lyot depolariser (DPU-25 Quartz-Wedge Achromatic Depolariser, uncoated 190--2500~nm, Thorlabs) was installed between the rotating polariser and the collimator lens to address the issues identified in Sec.~\ref{sec:results} below where an interaction between the rotating polarised beam output by the polariser and the dichroic mirror coatings is identified.  By depolarising the beam after the polariser (since at that stage we are only interested in measuring its time  variable intensity) we reduce this interaction. 

\item[\textbf{2014-06-08} (56816 MJD)] On this date the depolariser was moved from the input to the output of the collimator.  By moving it to the collimated beam, its depolarisation efficiency was increased since in a Lyot type depolariser the depolarisation obtained is related to the incident beam width. 

\item[\textbf{2015-06-27} (57200 MJD)] On this date the re-aluminisation of the primary and secondary mirrors was undertaken. 
The mirror recoating gave a large (factor $\sim 2$) throughput increase. 
Additionally, the polariser rotation was slowed down from one revolution per second to $\sim 0.4$ rotations per second with the change of motor gearbox ratio.  
This was done in order to improve signal-to-noise ratio for faint targets.  
A side effect of this change was that the direction of rotation was reversed. 
This caused the numerical sign of the Stokes u=U/I parameter to flip compared with data taken before this date. 
\end{description}

Hardware changes to the system introduced instrumental polarisation and depolarisation. Therefore we decided to split all the data
within respect to the dates of four hardware changes into five epochs. The knowledge of the 
system calibration in each of the epochs is required in terms of proper data analysis. 
Information about standard stars used in our analysis are gathered in Table~\ref{tab:polstan}
and Table~\ref{tab:polstan2} \citep{Turnshek1990, Schmidt1992}, regarding their photometric as well as polarimetric
characteristics, respectively.

\begin{table*}
 \centering
  \caption{Polarisation coordinates of standard stars, brightness and spectral types. Data analysis of four first listed standard stars 
  is presented in Sec.~\ref{sec:data}, while the results for all nine standard stars based on the data obtained during 
  the last (fifth) epoch are given in Sec.~\ref{sec:summary}.}
  \begin{tabular}{lcccccccccc}
  \hline
  \label{tab:polstan}
Name       & RA          & Dec          &  B    & V     & R     & I & Spec. Type & Comments\\
\hline
BD~+59~389 & 02 02 42.09 & +60 15 26.46 & 10.03 & 9.08  & 8.49  & 8.23  & F0Ib      & --- \\
BD~+64~106 & 00 57 36.71 & +64 51 26.50 & 10.87 & 10.29 & 9.90  & 9.84  & B1        & --- \\
G~191-B2B  & 05 05 30.61 & +52 49 51.96 & 11.43 & 11.67 & 11.82 & 11.45 & DA.8  & White Dwarf \\
HD~14069   & 02 16 45.19 & +07 41 10.67 & 9.21  & 9.06  & 8.97  & 8.91  & A0        & --- \\
\hline
BD~+25~727 & 04 44 24.92 & +25 31 42.72 & 10.20 & 9.55  & 9.11   & 8.98  & A2 III   & --- \\
HD~155528  & 17 12 19.95 & -04 24 09.26 & 10.05 & 9.61  & 9.32   & 9.16  & B9       & --- \\
HD~215806  & 22 46 40.24 & +58 17 43.94 & 9.55  & 9.21  & 8.99   & ---  & B0Ib     & --- \\
HILT~960   & 20 23 28.53 & +39 20 59.05 & 11.45 & 10.46 & 9.84   & 9.49  & B0V      & --- \\
VI~CYG~12  & 20 32 40.96 & +41 14 29.29 & 14.45 & 11.78 & 11.96  & 8.41  & B3-4 Ia+ & --- \\
\hline

\end{tabular}
\end{table*}

\begin{table}
 \centering
  \caption{Summary of the archival polarimetric measurements of standard stars, ref: H - \textit{The Hubble Space Telescope Northern-Hemisphere grid of stellar polarimetric standard stars} \citep{Schmidt1992}, S - \textit{Systematic variations in the wavelength dependence of interstellar linear polarisation} \citep{Whittet1992}.}
  \begin{tabular}{lcccc}
  \hline
  \label{tab:polstan2}
Name       & Johnson filter & PD [\%]         & PA [$^{\circ}$]           & ref\\
\hline
BD~+59~389 & B &  6.345 $\pm$ 0.035 & 98.14 $\pm$ 0.16   & H\\
           & V &  6.701 $\pm$ 0.015 & 98.09 $\pm$ 0.07   & H\\
           & R &  6.430 $\pm$ 0.022 & 98.14 $\pm$ 0.10   & H\\
           & I &  5.797 $\pm$ 0.023 & 98.26 $\pm$ 0.11   & H\\
BD~+64~106 & B & 5.506 $\pm$ 0.090 & 97.15 $\pm$ 0.47   & H\\
           & V & 5.687 $\pm$ 0.037 & 96.63 $\pm$ 0.18   & H\\
           & R & 5.150 $\pm$ 0.098 & 96.74 $\pm$ 0.54   & H\\
           & I &  4.696 $\pm$ 0.052 & 96.89 $\pm$ 0.32   & H\\
G~191-B2B  & B &  0.090 $\pm$ 0.048 & --- & H\\
           & V &  0.061 $\pm$ 0.038 & --- & H\\
HD~14069   & B &  0.111 $\pm$ 0.036 & --- & H\\
           & V &  0.022 $\pm$ 0.019 & --- & H\\
\hline
BD~+25~727 & B &  5.930 $\pm$ 0.070 & 31.00 $\pm$ 1.00   & S\\
           & V &  6.290 $\pm$ 0.050 & 32.00 $\pm$ 1.00   & S\\
           & R &  6.290 $\pm$ 0.070 & 31.00 $\pm$ 1.00   & S\\
           & I &  5.680 $\pm$ 0.070 & 31.00 $\pm$ 1.00   & S\\
HD~155528  & B &  4.612 $\pm$ 0.038 & 91.24 $\pm$ 0.24   & H\\
           & V &  4.986 $\pm$ 0.064 & 92.61 $\pm$ 0.37   & H\\
           & R &  ---               & ---                & -\\
           & I &  ---               & ---                & -\\
HD~215806  & B &  1.870 $\pm$ 0.040 & 66.00 $\pm$ 1.00   & S\\
           & V &  1.840 $\pm$ 0.050 & 67.00 $\pm$ 1.00   & S\\
           & R &  1.830 $\pm$ 0.040 & 66.00 $\pm$ 1.00   & S\\
           & I &  1.530 $\pm$ 0.050 & 67.00 $\pm$ 1.00   & S\\
HILT~960   & B &  5.720 $\pm$ 0.061 & 55.06 $\pm$ 0.31   & H\\
           & V &  5.663 $\pm$ 0.021 & 54.79 $\pm$ 0.11   & H\\
           & R &  5.210 $\pm$ 0.029 & 54.54 $\pm$ 0.16   & H\\
           & I &  4.455 $\pm$ 0.030 & 53.96 $\pm$ 0.19   & H\\
VI~CYG~12  & B &  9.670 $\pm$ 0.100 & 119.00 $\pm$ 1.00   & S\\
           & V &  8.947 $\pm$ 0.088 & 115.03 $\pm$ 0.28   & H\\
           & R &  7.893 $\pm$ 0.037 & 116.23 $\pm$ 0.14   & H\\
           & I &  7.060 $\pm$ 0.050 & 117.00 $\pm$ 1.00   & S\\
\hline
\end{tabular}
\end{table}

All three detectors of RINGO3 are the EMCCD.
The EMGAIN setting affects SNR in such a way that
a lower EMGAIN means a higher read noise and so worse SNR.
During the observations the EMGAIN was set to 100 for the whole
time for BD~+64~106 and G~191-B2B, whereas for BD~+59~389 as well as for
HD~14069 the gain was set to 100 before 57200 MJD and was changed to
20 after this date, because both stars are brighter than 10th magnitude.
However for such bright standards, the read noise is not really important.

\section{Data analysis}
\label{sec:data}

Data analysis was conducted in the following order: data gathering, image stacking, measuring flux, Stokes parameters
calculations, PD and PA derivation with appropriate polarimetric corrections. Each step of analysis is described in detail below.

Firstly, we downloaded over 100,000 data frames from the public
archive, including
73,552 frames for polarised sources
(9,888 for BD~+59~389; 10,536 for BD~+64~106; 14,872 for BD~+25~727; 17,280 for VI~Cyg~12; 17,304 for Hilt~960; 
1,440 for CRL~2688; 1,080 for HD~155528; 1,152 for HD~215806)
and 30,672 frames for zero polarised sources (21,352 for G~191-B2B and 9,320 for HD~14069).
Data analysis as well as the results of CRL~2688 are the topic of separated
study, because it is extended source and not the point source as all the other
targets. 
For the first part of the data analysis (Sec.~\ref{sec:data}
and Sec.~\ref{sec:results}) we concentrated only on four sources,
two polarised ones, i.e. BD~+59~389 and BD~+64~106, and two zero-polarised, i.e. G~191-B2B and HD~14069. 
However, all sources listed in Table~\ref{tab:polstan} were used for the summary and conclusions (Sec.~\ref{sec:summary}) and their results are presented in Table~\ref{tab:factors5} as well as in Fig.~\ref{fig:lt_vs_robopol}.

The data were debiased, flat fielded and had a World Coordinate System (WCS) fitted by 
the standard pipeline running at the telescope.
Using the WCS coordinates we identified our targets and
with the Source-Extractor \citep{Bertin1996} we extracted the target flux
and its corresponding error with the best aperture.
Thus, our input data consist of the flux value, its error as well as
an information at which rotation angle (one of 0$^{\circ}$, 45$^{\circ}$,
90$^{\circ}$, 135$^{\circ}$, 180$^{\circ}$, 225$^{\circ}$ 270$^{\circ}$, and
315$^{\circ}$) this value was measured. 
Counting that the single set of observation consists of 8-frames
there were 1236, 1317, 2669 and 1165 polarimetric measurements
for BD~+59~389 and BD~+64~106, G~191-B2B and HD~14069, respectively.
We selected only such observations for which all 8 flux values were
available after the basic data reduction. 
This constrain removed 43, 50, 221 and 163 data points from sets of 8-frames
for BD~+59~389 and BD~+64~106, G~191-B2B and HD~14069, respectively.
Additionally, we applied the constraint on the PD error lower than $1\%$
to data of all sources and the PD of G~191-B2B to be lower than $3\%$.
We also filtered out the observations with seeing greater than 5 arcsec.
These conditions additionally removed the following number of data points for each source:
12 points from BD~+59~389 data,
21 points from BD~+64~106 data,
121 points from G~191-B2B data and
19 points from HD~14069 data.
As a result we obtained a total of 5737
polarimetric measurements, i.e. 1181, 1246, 2327
and 983 for BD~+59~389, BD~+64~106, G~191-B2B and HD~14069, respectively. 

To calculate the Stokes I, Q, and U we used method described by \cite{Sparks1999}
for n--polarisers. This method was very successfully used before. One of
the most extreme cases of using the n-polarisers method is the case of the Crab
pulsar presented by \cite{Slowikowska2009} where measurements through as many as 180
positions of the rotating polariser were used to calculate PA and
PD as a function of rotational phase of the neutron star. Regarding
stellar optical polarimetry (apart form the Sun) this results gave the highest
time resolution so far achieved, i.e. the polarisation was measured in the time
scales down to ten microseconds. The greatest advantage of using more than three 
polarisers is the significant reduction of the errors of PD and PA.

For each source we calculated normalised Stokes parameters q=Q/I and u=U/I
for data acquired from all three cameras. Because the RINGO3 hardware was changed
multiple times (see Sec.~\ref{sec:obs})
during the time span that covers our whole data set,
we had to split the data into five parts 
and analyse them separately to remove the influence of the system
changes. The resulting q-u diagrams are shown in Figs.~
\ref{fig:bd59_shift}, \ref{fig:bd64_shift}, \ref{fig:g191_shift} and \ref{fig:hd14_shift}
for BD~+59~389, BD~+64~106, G~191-B2B and HD~14069, respectively.
Each q-u diagram consists of 15 panels. There are three rows corresponding to
three camera colours (blue, green and red from the top to the bottom) and five columns for
five MJD epochs. For polarised sources the q-u diagrams show points aligned in circular shape.
This is caused by the fact that the LT telescope has an alt-az mount, therefore
the data need to be corrected for the sky position angle of the image, i.e.
how the rotation of the image relates to the North (see Eq.~\ref{eq:truePA}).
The radius of the circle corresponds to the measured PD.
Significant instrumental polarisation causes the centres of the circles to be shifted away
from the zero-zero point. The same behaviour is visible in the q-u diagrams
of zero-polarised stars. However, in such cases, instead of circles the data points
form clumps. In order to remove the instrumental polarisation one needs to
shift the data points to zero-zero origin. In case of polarised standard stars we fit
the circles to the data points to get the coordinates of the circles centres,
while in case of zero-polarised sources we calculated the weighted means
in both directions (q, u) and treated these as the centres of the clumps. 
These way we obtained the shifts' values from the zero-zero point 
for four sources in three colours in five epochs. 
These shifts should be equal to the instrumental polarisation.
It can be different for each camera, but should be the same for all sources
in case of one camera. All the centres of circles and clumps for three
colours and five epochs are presented in Fig.~\ref{fig:shifts_plot}.
There is a significant scatter of the centres coordinates in the  
first two epochs, that cover the time span from 56200 MJD to 56638 MJD.
However, in the next epochs the system is getting more stable in terms
of instrumental polarisation. The centres have similar coordinates for
each source which means that they translate to the similar instrumental 
polarisation values. This way one is able to remove the instrumental
contribution to the measured PD. 

For each single observation we calculated and corrected Stokes parameters 
for the instrumental polarisation. In the next step we calculated the
polarisation degree (PD) and the position angle (PA).
For each MJD range between LT hardware changes
the instrumental polarisation,
PD factors and PA shifts as defined below:
\begin{equation}
\mathrm{true~PD = \frac{measured~PD}{PD~factor}}
\label{eq:truePD}
\end{equation}
and
\begin{equation}
\mathrm{true~PA = measured~PA + ROTSKYPA + PA~shift}
\label{eq:truePA}
\end{equation}
were calculated, where ROTSKYPA is the sky position angle of the image.

To derive the factors and shifts we scaled the measured PD and PA with respect to published polarisation
results of polarisation standard stars \citep{Schmidt1992}. In case of the data obtained with the blue 
camera we scale accordingly to the published averaged values for the B and V Johnson filters
(see Table~\ref{tab:polstan2}), i.e. to 6.523\%, 5.5965\% for BD~+59~389 and BD~+64~106, respectively. 
Similarly to the PD factors, the PA shifts for the blue camera were
calculated with respect to the averaged values of the PA in the B and V Johnston filters, 
i.e. $98.115^{\circ}$ and $96.89^{\circ}$ for BD~+59~389 and BD~+64~106, respectively.
For all data scaling we used averaged PD and PA factors of polarised standard stars, i.e. BD~+59~389 and BD~+64~106. 

PD of zero-polarised standard stars calculated from shifted q and u values is shown in 
Fig.~\ref{fig:g191_pd}, \ref{fig:hd14_pd} for G~191-B2B and HD~14069, respectively.
The data form the red camera have much bigger scatter than the data from the blue
camera in both cases.
Obtained PD, according to the Eq.~\ref{eq:truePD}, and PA,
acoording to Eq.~\ref{eq:truePA}, for two non-zero polarised standard stars are shown in 
Fig.~\ref{fig:bd59_pd}, \ref{fig:bd64_pd} and in Fig.~\ref{fig:bd59_pa}, \ref{fig:bd64_pa} 
for BD~+59~389 and BD~+64~106, respectively. 
In all cases it is clearly visible that the single data points in the last epoch have
the smallest uncertainties, as well as they are less scattered in comparison to the other
epochs. However, in case of BD~64+106 a quite strong variability also is present. 
We will be able to have stronger conclusions on the variability of this
source from the data obtained after the latest hardware change.

\section{Results}
\label{sec:results}

The resulting PD factors and PA shifts for five consecutive epochs are gathered in Table~\ref{tab:factors}. The average of averaged PD factors for all five epochs
(56200--57400 MJD) is on the level of 0.802 for all three colours, while the averaged
PA shift is around 76.297 degrees for the first three
epochs (56200--56816 MJD), \ensuremath{-41.024} degrees for the fourth epoch (56816--57200 MJD) and \ensuremath{-74.886} degrees for the last epoch (57200--57400 MJD), again for all three colours.

\begin{knitrout}
\definecolor{shadecolor}{rgb}{0.969, 0.969, 0.969}\color{fgcolor}\begin{figure}
\includegraphics[width=\maxwidth]{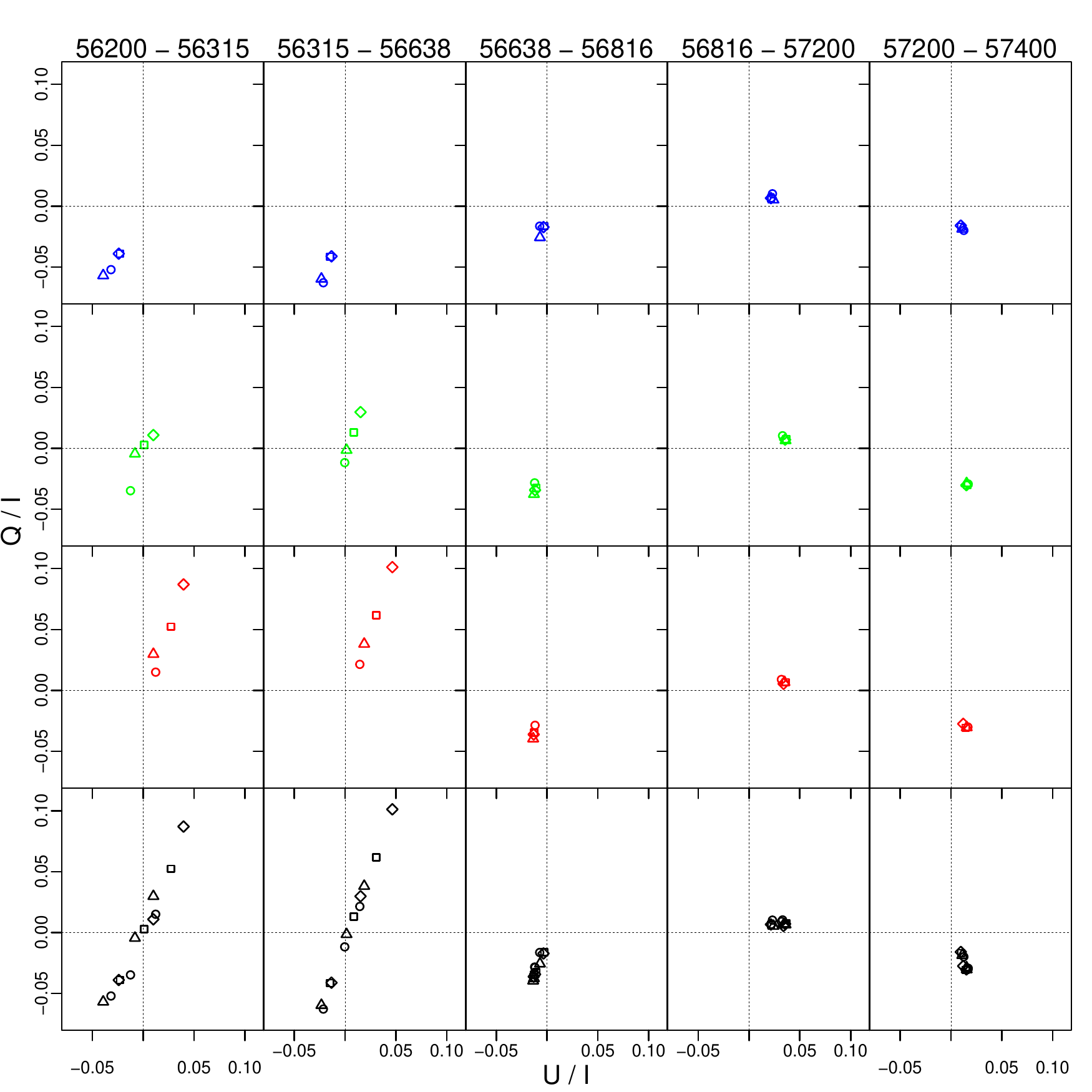} \caption[Weighted means of the normalised Stokes Q/I and U/I parameters for the unpolarised standard stars 
 G~191-B2B (rhombus) and HD~14069 (square) and the centres of fitted circles for polarised standard stars 
 BD~+59~389 (circle) and  BD~+64~106 (triangle) as a function of time (five epochs in MJD)]{Weighted means of the normalised Stokes Q/I and U/I parameters for the unpolarised standard stars 
 G~191-B2B (rhombus) and HD~14069 (square) and the centres of fitted circles for polarised standard stars 
 BD~+59~389 (circle) and  BD~+64~106 (triangle) as a function of time (five epochs in MJD). The bottom row contains all the 
 points in all colours for all the standard stars for each epoch. The scatter 
 is highest in the first two epochs (56200--56638 MJD) with the lowest value of scatter in the case of 
 the blue camera and the biggest for the red one. In the last two epochs of 56816--57200 MJD and 
 57200--57400 MJD all the points are gathered around the same point which is very close to zero--zero. 
 This shows that the hardware changes improved the system significantly.}\label{fig:shifts_plot}
\end{figure}

\end{knitrout}

As a test we also calculated the PA and PD with their errors for the case of 3
polarisers (0$^{\circ}$, 45$^{\circ}$, 135$^{\circ}$) for the non-zero polarised source with the highest
number of measurements, i.e. BD~+59~389. The results show that using 8 polarisers
reduces the errors of PA and PD by a mean factor of $\sim 1.95$ (with the maximum
of 4.15) with respect to the case of 3 polarisers. 
Using more than 3 polarisers ensures diminishing of error estimations of
the Stokes parameters and therefore PD and PA errors.

\begin{table*}
\centering
\caption{Polarisation PD factors as well as PA shift for five epochs. True PD is calculated as measured PD divided by PD factor, and true PA is calculated as PA shift added to measured PA.}
\label{tab:factors}
\begin{tabular}{c|llll|rrrr}
  \hline
    \multirow{2}{*}{MJD range} & \multicolumn{4}{c|}{PD factor} & \multicolumn{4}{c|}{PA shift [$^\circ$]}\\
        & blue & green & red & average & blue & green & red & average\\
  \hline
    56200 - 56315 
        & 0.816 
        & 0.919$^a$ 
        & 0.808 
        & 0.812$^b$ 
        & 75.322 
        & 74.049 
        & 76.635 
        & 75.335 \\
    56315 - 56638 
        & 0.801 
        & 0.806 
        & 0.806 
        & 0.804 
        & 77.107 
        & 76.358 
        & 77.115 
        & 76.86 \\
    56638 - 56816 
        & 0.818 
        & 0.82 
        & 0.777 
        & 0.805 
        & 77.18 
        & 76.434 
        & 76.47 
        & 76.695 \\
    56816 - 57200 
        & 0.817 
        & 0.828 
        & 0.802 
        & 0.816 
        & \ensuremath{-40.813} 
        & \ensuremath{-41.38} 
        & \ensuremath{-40.88} 
        & \ensuremath{-41.024} \\
    57200 - 57400 
        & 0.754 
        & 0.763 
        & 0.808 
        & 0.775 
        & \ensuremath{-74.749} 
        & \ensuremath{-75.314} 
        & \ensuremath{-74.594} 
        & \ensuremath{-74.886} \\
\hline

\multicolumn{9}{l}{$^a$ \tiny{There are not many data for the green camera during the first epoch, therefore the calculated PD factor is not meaningful.}}\\
\multicolumn{9}{l}{$^b$ \tiny{This average is calculated only from the PD factors of the blue and the red cameras.}}
\end{tabular}
\end{table*}

\subsection{Time series}
We also show normalised Stokes parameters (Q/I and U/I) of two unpolarised standard stars (G~191-B2B and HD~14069) as a 
function of time for five consecutive epochs (Fig.~\ref{fig:time_series_g191_qi}, \ref{fig:time_series_hd14_qi}, 
\ref{fig:time_series_g191_ui} and \ref{fig:time_series_hd14_ui}). For unpolarised standard, that by definition 
should not change as a function of time, one expects that Stokes parameters will be constant in time. From obtained 
plots one can read that this condition is the best fulfilled for G~191-B2B standard in the last epoch. 
HD~14069 shows some small, but significant trend in the fourth and fifth epoch, when the instrument was 
already well calibrated, especially in the case of the Stokes parameter U/I. 
For each standard, colour and epoch the linear model was fitted and the obtained $a$ (slope) and $b$ (intercept) parameters 
are gathered in the Table~\ref{tab:qi_fits} and \ref{tab:ui_fits}. These models can be used for the data 
calibration with respect to each epoch and colour. The most often observed unpolarised standard was G~191-B2B. 
It is the most stable over time as well, therefore we recommend this target as the best calibration target 
for other sources. In case of HD~14069 the small changes in U/I Stokes parameter, observed even in the last 
epoch, might be intrinsic and not instrumental.

\begin{knitrout}
\definecolor{shadecolor}{rgb}{0.969, 0.969, 0.969}\color{fgcolor}\begin{figure}
\includegraphics[width=\maxwidth]{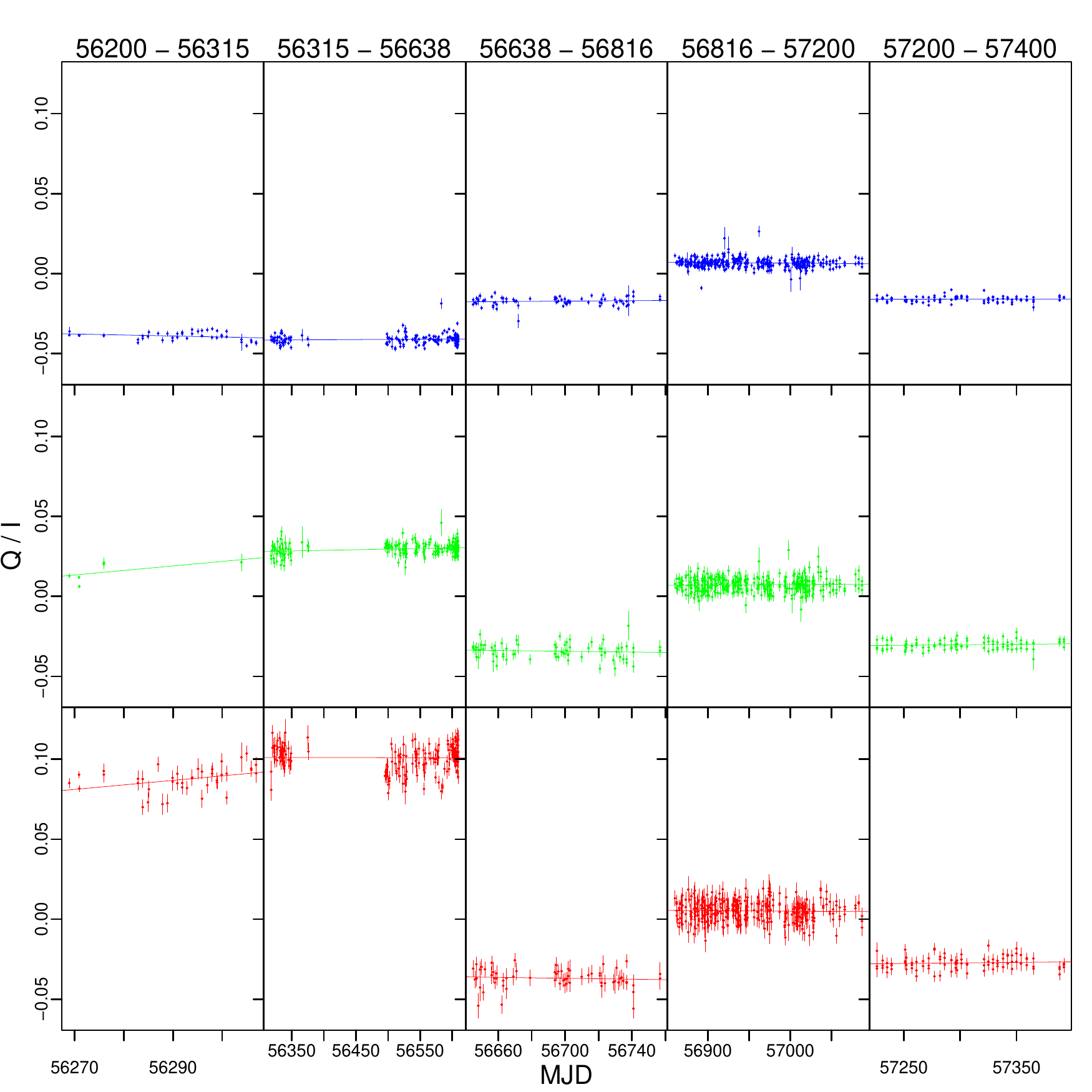} \caption[Normalised Stokes parameter Q/I for G~191-B2B as a function of time]{Normalised Stokes parameter Q/I for G~191-B2B as a function of time. There are not many data points from the green camera in the first epoch, therefore the Q/I dependence on time is not well constrained.}\label{fig:time_series_g191_qi}
\end{figure}

\end{knitrout}

\begin{knitrout}
\definecolor{shadecolor}{rgb}{0.969, 0.969, 0.969}\color{fgcolor}\begin{figure}
\includegraphics[width=\maxwidth]{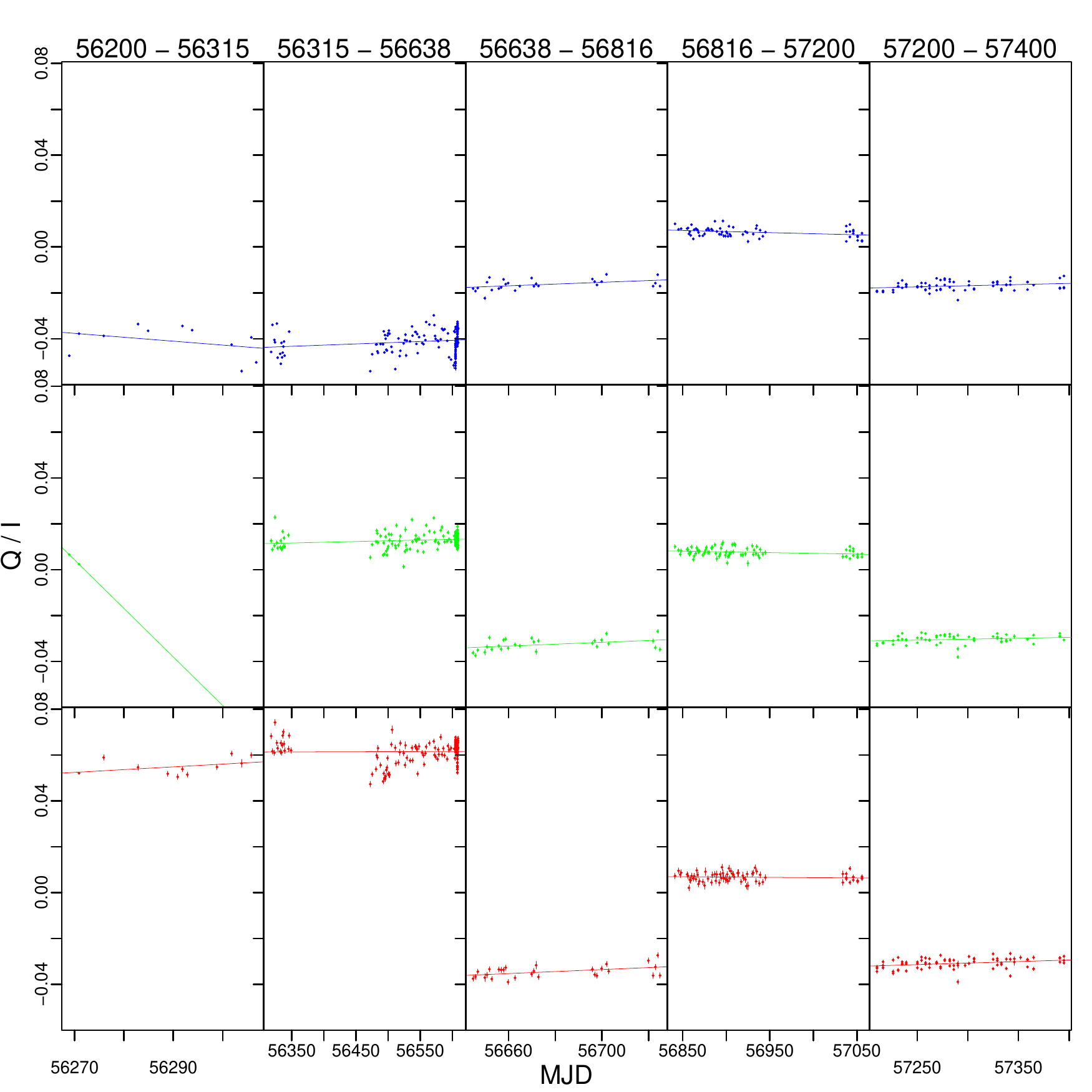} \caption[Normalised Stokes parameter Q/I for HD~14069 as a function of time]{Normalised Stokes parameter Q/I for HD~14069 as a function of time. There are not many data points from the green camera in the first epoch, therefore the Q/I dependence on time is not well constrained.}\label{fig:time_series_hd14_qi}
\end{figure}

\end{knitrout}

\begin{knitrout}
\definecolor{shadecolor}{rgb}{0.969, 0.969, 0.969}\color{fgcolor}\begin{figure}
\includegraphics[width=\maxwidth]{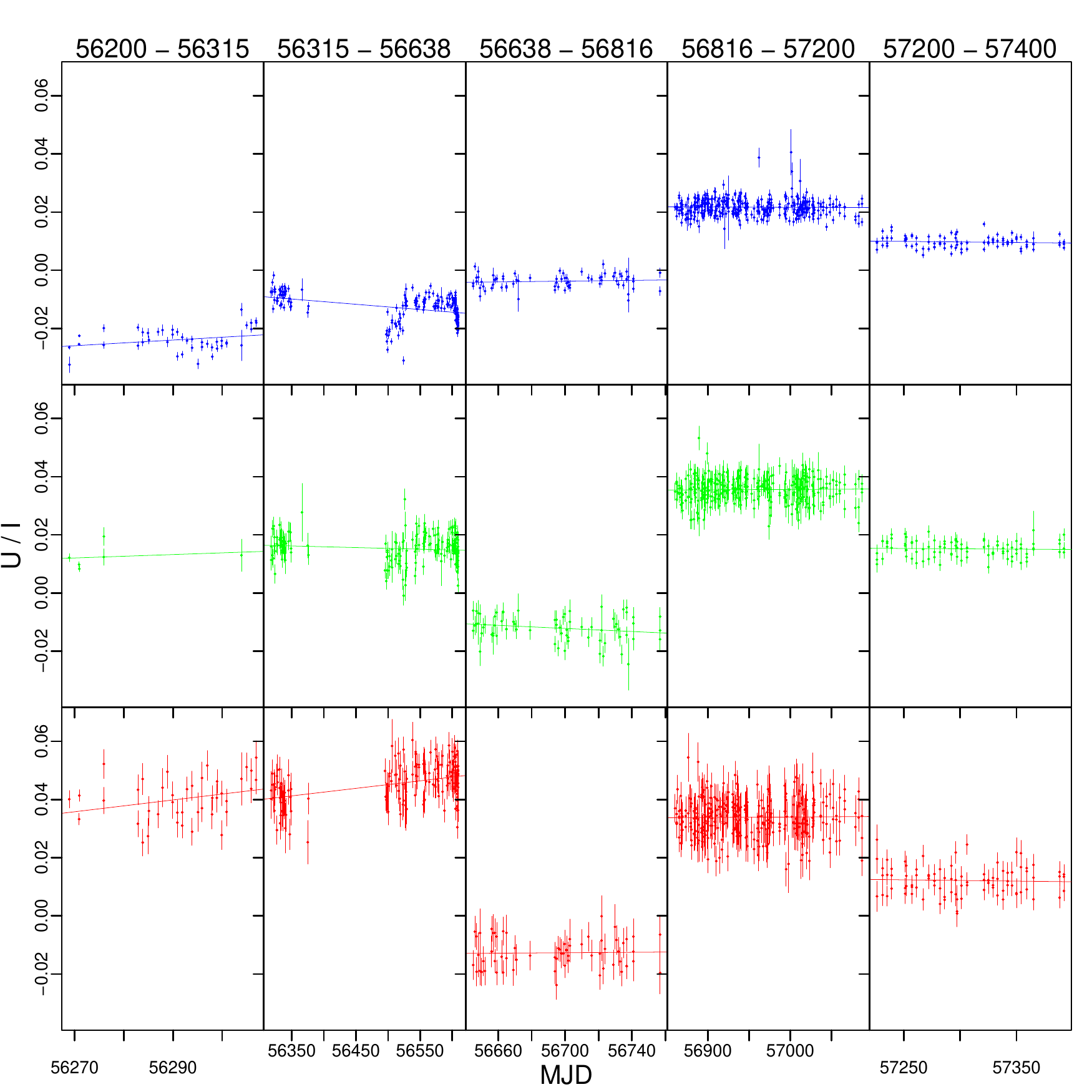} \caption[Normalised Stokes parameter U/I for G~191-B2B as a function of time]{Normalised Stokes parameter U/I for G~191-B2B as a function of time. There are not many data points from the green camera in the first epoch, therefore the U/I dependence on time is not well constrained.}\label{fig:time_series_g191_ui}
\end{figure}

\end{knitrout}

\begin{knitrout}
\definecolor{shadecolor}{rgb}{0.969, 0.969, 0.969}\color{fgcolor}\begin{figure}
\includegraphics[width=\maxwidth]{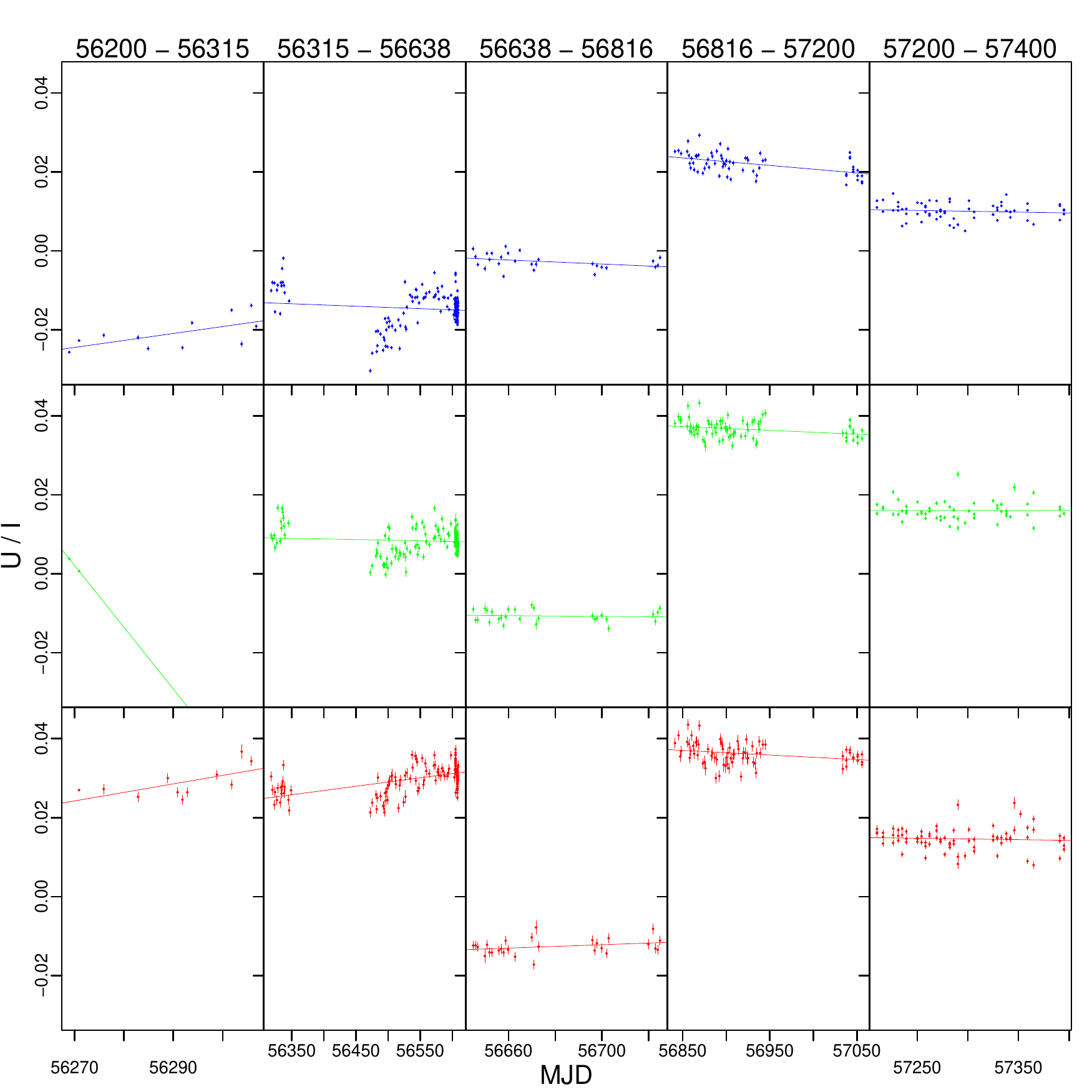} \caption[Normalised Stokes parameter U/I for HD~14069 as a function of time]{Normalised Stokes parameter U/I for HD~14069 as a function of time. There are not many data points from the green camera in the first epoch, therefore the U/I dependence on time is not well constrained.}\label{fig:time_series_hd14_ui}
\end{figure}

\end{knitrout}

For the fourth and fifth epochs the weighted means of Q/I and U/I along with the weighted standard deviations 
for G~191-B2B and HD~14069 were calculated and are gathered in Table~\ref{tab:qi_ui_comp_4} and \ref{tab:qi_ui_comp_5}.

\begin{table}
\caption{Comparison of Q/I and U/I for two zero-polarised standard stars in the fourth epoch 56816--57200 MJD.}
\label{tab:qi_ui_comp_4}
\centering
\begin{tabular}{lcc}
    \hline
       & G~191-B2B & HD~14069\\
    \hline
    Q/I$_\mathrm{blue}$ 
    &0.00664 $\pm$ 0.00007
    &0.00631 $\pm$ 0.00006\\
    
    Q/I$_\mathrm{green}$
    &0.00722 $\pm$ 0.00017
    &0.00759 $\pm$ 0.0001\\
    
    Q/I$_\mathrm{red}$  
    &0.00538 $\pm$ 0.00027
    &0.00667 $\pm$ 0.00014\\
    
    &&\\
    
    U/I$_\mathrm{blue}$ 
    &0.02152 $\pm$ 0.00007
    &0.02172 $\pm$ 0.00006\\
    U/I$_\mathrm{green}$
    &0.03561 $\pm$ 0.00017
    &0.03642 $\pm$ 0.0001\\
    U/I$_\mathrm{red}$  
    &0.03396 $\pm$ 0.00027
    &0.03606 $\pm$ 0.00014\\
    \hline
\end{tabular}
\end{table}

\begin{table}
\caption{Comparison of Q/I and U/I for two zero-polarised standard stars in the fifth epoch 57200--57400 MJD.}
\label{tab:qi_ui_comp_5}
\centering
\begin{tabular}{lcc}
    \hline
       & G~191-B2B & HD~14069\\
    \hline
    Q/I$_\mathrm{blue}$ 
    &\ensuremath{-0.01593} $\pm$ 0.00009
    &\ensuremath{-0.01683} $\pm$ 0.00003\\
    
    Q/I$_\mathrm{green}$
    &\ensuremath{-0.03023} $\pm$ 0.00022
    &\ensuremath{-0.03013} $\pm$ 0.00006\\
    
    Q/I$_\mathrm{red}$  
    &\ensuremath{-0.0274} $\pm$ 0.00037
    &\ensuremath{-0.03071} $\pm$ 0.00008\\
    
    &&\\
    
    U/I$_\mathrm{blue}$ 
    &0.0097 $\pm$ 0.00009
    &0.01011 $\pm$ 0.00003\\
    U/I$_\mathrm{green}$
    &0.01519 $\pm$ 0.00022
    &0.01588 $\pm$ 0.00006\\
    U/I$_\mathrm{red}$  
    &0.01207 $\pm$ 0.00037
    &0.01446 $\pm$ 0.00008\\
    \hline
\end{tabular}
\end{table}

\begin{table}
\caption{The PD factors and the PA shifts in the fourth epoch 56816--57200 MJD.}
\label{tab:factors_shifts_4}
\centering
\begin{tabular}{lllll}
\hline
colour & PD factor & PA shift [deg] & PD factor & PA shift [deg]\\
\hline
&\multicolumn{2}{c}{BD~+59~389} &\multicolumn{2}{c}{BD~+64~106}\\
\hline
    blue 
        & 0.767 
        & \ensuremath{-39.888}
        & 0.868 
        & \ensuremath{-41.738}\\
    green 
        & 0.768 
        & \ensuremath{-40.571}
        & 0.887 
        & \ensuremath{-42.189}\\
    red 
        & 0.75 
        & \ensuremath{-40.163}
        & 0.853 
        & \ensuremath{-41.596}\\
\hline
\end{tabular}
\end{table}

\begin{table}
\caption{The PD factors and the PA shifts in the fifth epoch 57200--57400 MJD.}
\label{tab:factors_shifts_5}
\centering
\begin{tabular}{lllll}
\hline
colour & PD factor & PA shift [deg] & PD factor & PA shift [deg]\\
\hline
&\multicolumn{2}{c}{BD~+59~389} &\multicolumn{2}{c}{BD~+64~106}\\
\hline
    blue 
        & 0.793 
        & \ensuremath{-76.055}
        & 0.715 
        & \ensuremath{-73.443}\\
    green 
        & 0.773 
        & \ensuremath{-76.424}
        & 0.754 
        & \ensuremath{-74.205}\\
    red 
        & 0.809 
        & \ensuremath{-75.552}
        & 0.807 
        & \ensuremath{-73.635}\\
\hline
\end{tabular}
\end{table}

\subsection{Correlations}
We also checked if there are any correlations between the observing conditions and derived values. 
There are no correlations of the PD or the PA with the Moon phase or the Moon distance from the 
source in any of the observed energy ranges. Pearson's correlation coefficients for all
sources in all energy ranges were $|r| < 0.1$. An example plot for the case of
BD~+59~389 is shown in Fig.~\ref{fig:cor_moon}. 

However, there seem to be weak positive correlations between PD and seeing in BD +59 389. 
The larger the seeing, the larger the PD. The correlation coefficients are
0.47, 0.45 
and 0.49
for blue, green and red cameras, respectively. In this case, the most 
likely explanation is that there is contamination from a nearby star which is 16 arcseconds 
away from the target. No significant correlations were detected between PA and seeing. 
Figure~\ref{fig:cor_seeing} shows the PD as a function of seeing, measured as the FWHM of 
the radial light distribution of the star on the detector. BD~+59~389 is the only star 
that shows this weak relationship. As a comparison, we also present data for BD~+64~106 
(lower panel of Fig.~\ref{fig:cor_seeing}). No correlation is seen in this case.

\begin{knitrout}
\definecolor{shadecolor}{rgb}{0.969, 0.969, 0.969}\color{fgcolor}\begin{figure}
\includegraphics[width=\maxwidth]{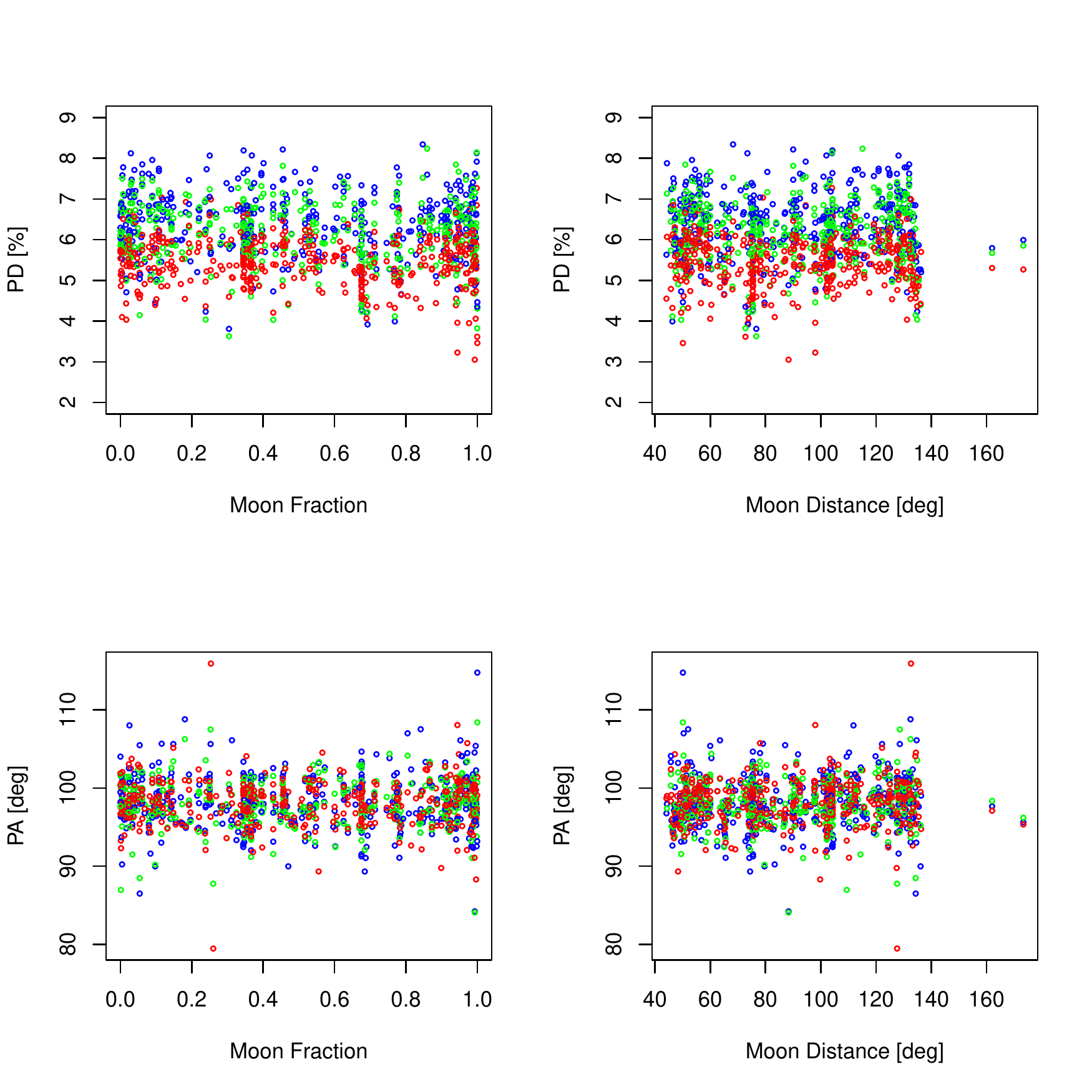} \caption[The PD and the PA of BD~+59~389 as a function of the Moon fraction and the Moon distance from the source]{The PD and the PA of BD~+59~389 as a function of the Moon fraction and the Moon distance from the source. There are no correlations between these values. Colours correspond to the data obtained with the respective cameras.}\label{fig:cor_moon}
\end{figure}

\end{knitrout}

\begin{knitrout}
\definecolor{shadecolor}{rgb}{0.969, 0.969, 0.969}\color{fgcolor}\begin{figure}
\includegraphics[width=\maxwidth]{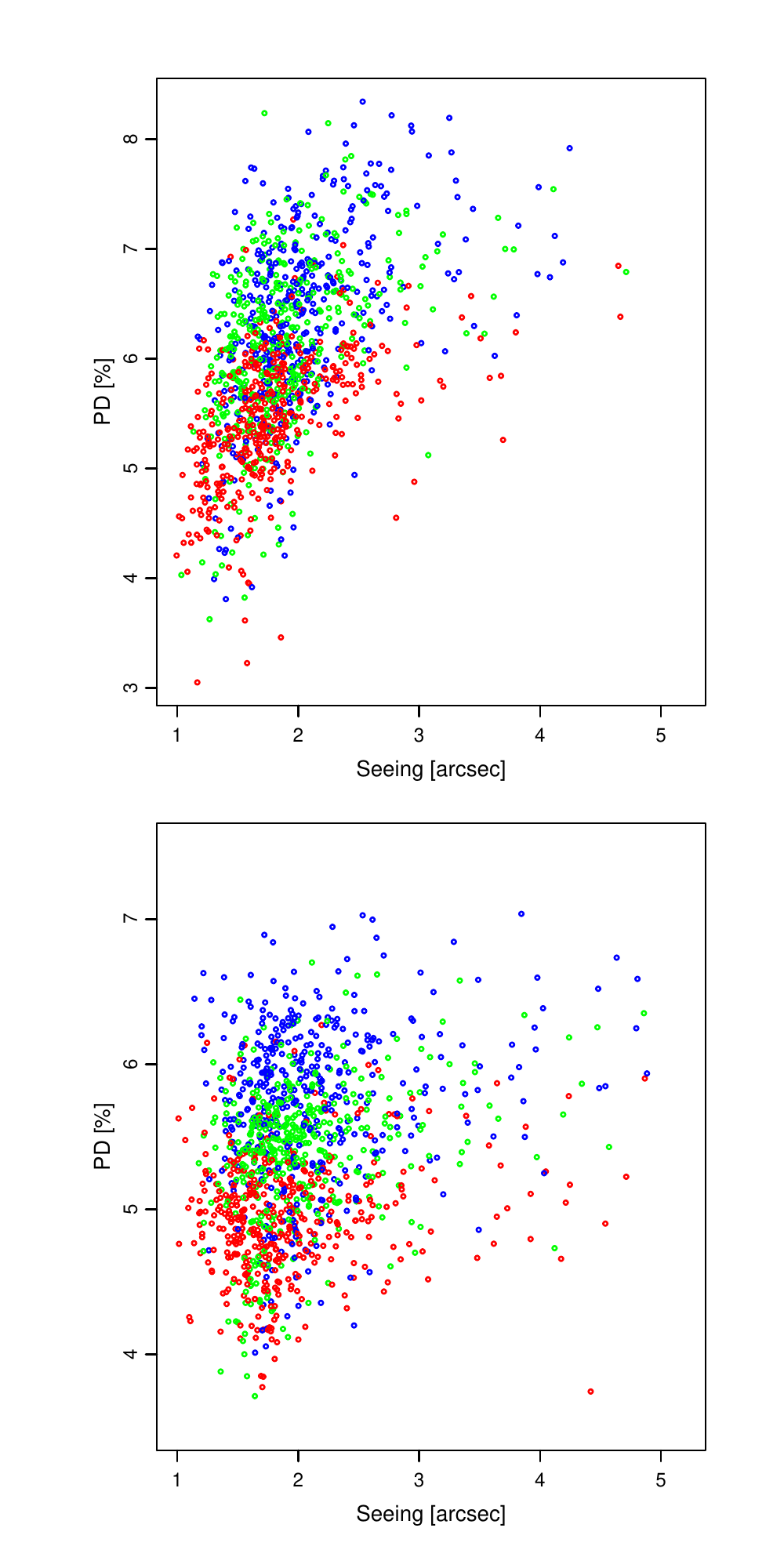} \caption[Dependence of the PD on the seeing for BD~+59~389 (upper panel) and BD~+64~106 (lower panel) in 
 three camera colours for all available measurements]{Dependence of the PD on the seeing for BD~+59~389 (upper panel) and BD~+64~106 (lower panel) in 
 three camera colours for all available measurements. BD~+59~389 is the only star on 
 our list that shows a weak dependence of PD on seeing.}\label{fig:cor_seeing}
\end{figure}

\end{knitrout}

\section{Summary and conclusions}
\label{sec:summary}

\begin{knitrout}
\definecolor{shadecolor}{rgb}{0.969, 0.969, 0.969}\color{fgcolor}\begin{figure}
\includegraphics[width=\maxwidth]{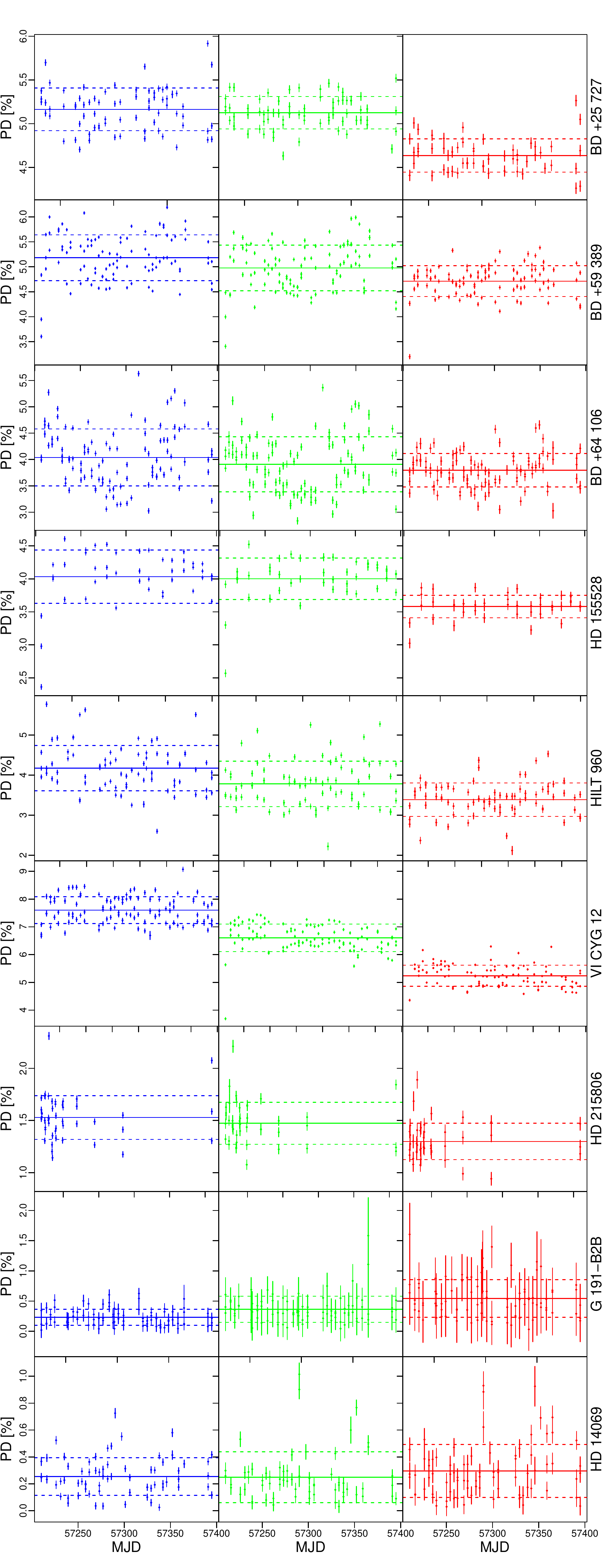} \caption{Measured PD not corrected for the PD factor of nine 
 calibration standard stars in three colours in the last MJD epoch.
 Solid lines correspond to the mean values, while dotted lines to one
 standard deviation in each single panel (see also Table~\ref{tab:factors5}).}\label{fig:pd9}
\end{figure}

\end{knitrout}

\begin{knitrout}
\definecolor{shadecolor}{rgb}{0.969, 0.969, 0.969}\color{fgcolor}\begin{figure}
\includegraphics[width=\maxwidth]{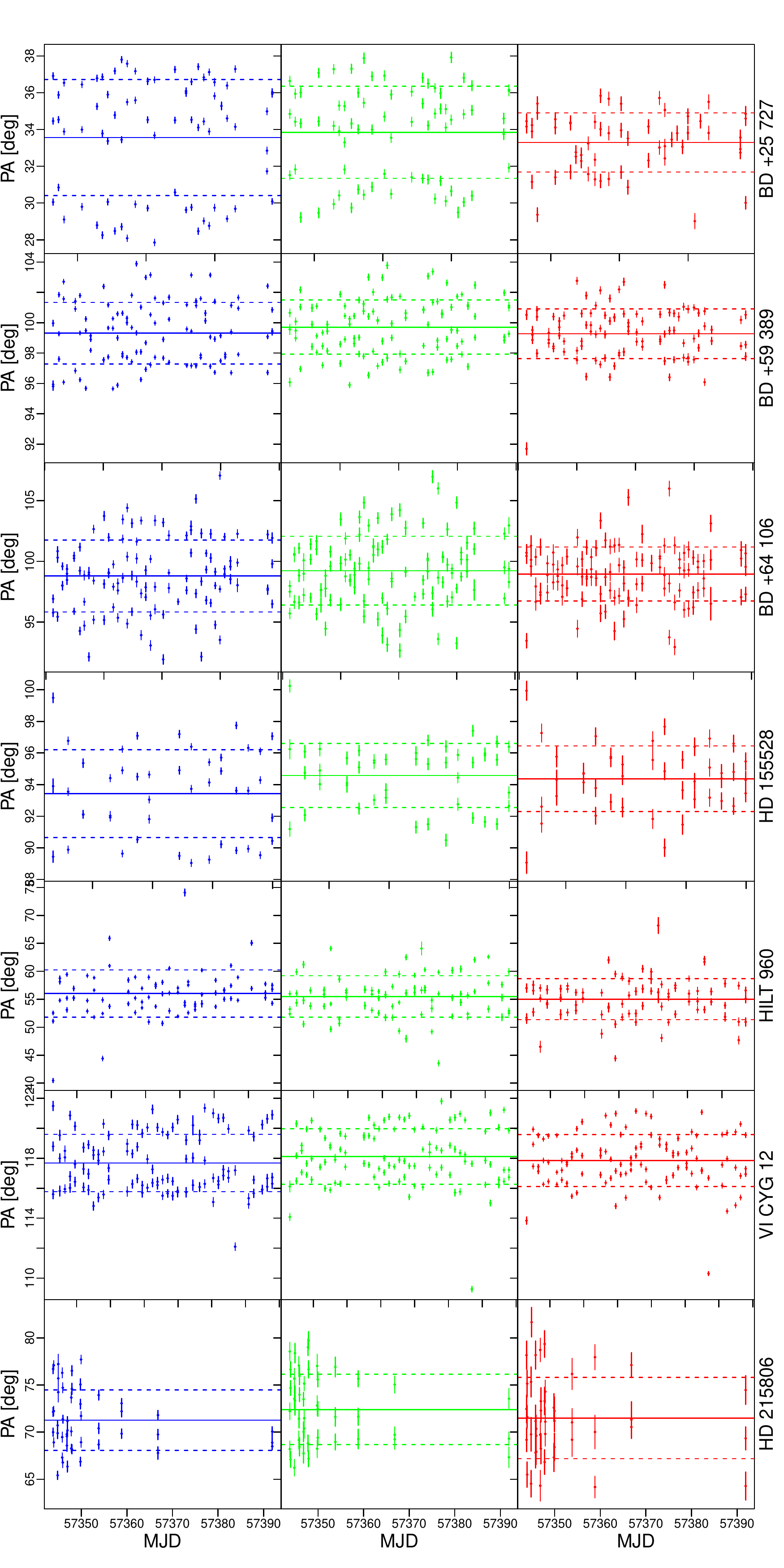} \caption{Measured PA of seven high polarisations standard stars in three
 colours in the last epoch (57200--57400 MJD). PA is corrected for the PA shift that is 
 taken as an averaged value from Table~\ref{tab:factors} for the fifth epoch, i.e. -74.9 degrees. 
 Solid lines correspond to the mean values, while dotted lines to one
 standard deviation in each single panel (see also Table~\ref{tab:factors5}).}\label{fig:pa9}
\end{figure}

\end{knitrout}

% latex table generated in R 3.2.2 by xtable 1.7-4 package
% Thu Feb  4 15:37:40 2016
\begin{table*}
\centering
\caption{The RINGO3 measured PD and PA values of polarised standard stars for the fifth epoch (57200--57400 MJD) together with their standard deviations ($\sigma_\mathrm{PD}$, $\sigma_\mathrm{PA}$) as well as with the respective catalogued values. Measured PD was not corrected for the PD factor, while the PA is corrected for the PA shift that is taken as an averaged value from Table~\ref{tab:factors} for the fifth epoch, i.e. -74.9 degrees. The catalogued values of PD, PA and magnitude for the blue camera are the mean values obtained for the B and V Johnston filters (see Table~\ref{tab:polstan2}), whereas green and red cameras correspond to the R and I, respectively. The Count column gives the number of data points in each colour for each source. The value of EMGAIN setting during observations is also given.} 
\label{tab:factors5}
\begin{tabular}{llrrrrrrrrr}
  \hline
Source & Camera & Count & PD [\%] & $\sigma_\mathrm{PD}$ [\%] & PD cat [\%] & PA [deg] & $\sigma_\mathrm{PA}$ [deg] & PA cat [deg] & Brightness [mag] & EMGAIN \\ 
  \hline
BD +25 727 & blue & 69 & 5.16 & 0.25 & 6.11 & 33.5 & 3.2 & 31.5 & 9.88 & 20 \\ 
  BD +25 727 & green & 68 & 5.12 & 0.19 & 6.29 & 33.8 & 2.5 & 31.0 & 9.11 & 20 \\ 
  BD +25 727 & red & 52 & 4.64 & 0.19 & 5.68 & 33.3 & 1.5 & 31.0 & 8.98 & 20 \\ 
  BD +59 389 & blue & 99 & 5.18 & 0.46 & 6.52 & 99.3 & 2.0 & 98.1 & 9.55 & 20 \\ 
  BD +59 389 & green & 101 & 4.98 & 0.46 & 6.43 & 99.7 & 1.8 & 98.1 & 8.49 & 20 \\ 
  BD +59 389 & red & 97 & 4.71 & 0.31 & 5.80 & 99.3 & 1.6 & 98.3 & 8.23 & 20 \\ 
  BD +64 106 & blue & 106 & 4.01 & 0.51 & 5.60 & 98.8 & 3.0 & 96.9 & 10.58 & 100 \\ 
  BD +64 106 & green & 107 & 3.89 & 0.51 & 5.15 & 99.3 & 2.8 & 96.7 & 9.90 & 100 \\ 
  BD +64 106 & red & 108 & 3.80 & 0.32 & 4.70 & 98.9 & 2.2 & 96.9 & 9.84 & 100 \\ 
  G 191-B2B & blue & 93 & 0.23 & 0.13 & 0.08 & $-$ & $-$ & $-$ & 11.55 & 100 \\ 
  G 191-B2B & green & 93 & 0.35 & 0.18 & $-$ & $-$ & $-$ & $-$ & 11.82 & 100 \\ 
  G 191-B2B & red & 92 & 0.54 & 0.31 & $-$ & $-$ & $-$ & $-$ & 11.45 & 100 \\ 
  HD 14069 & blue & 68 & 0.25 & 0.13 & 0.07 & $-$ & $-$ & $-$ & 9.13 & 20 \\ 
  HD 14069 & green & 60 & 0.25 & 0.18 & $-$ & $-$ & $-$ & $-$ & 8.97 & 20 \\ 
  HD 14069 & red & 81 & 0.30 & 0.20 & $-$ & $-$ & $-$ & $-$ & 8.91 & 20 \\ 
  HD 155528 & blue & 45 & 4.03 & 0.40 & 4.80 & 93.4 & 2.8 & 91.9 & 9.83 & 20 \\ 
  HD 155528 & green & 45 & 4.00 & 0.31 & $-$ & 94.6 & 2.0 & $-$ & 9.32 & 20 \\ 
  HD 155528 & red & 44 & 3.58 & 0.17 & $-$ & 94.4 & 2.1 & $-$ & 9.16 & 20 \\ 
  HD 215806 & blue & 48 & 1.53 & 0.21 & 1.85 & 71.3 & 3.2 & 66.5 & 9.38 & 20 \\ 
  HD 215806 & green & 47 & 1.47 & 0.20 & 1.83 & 72.4 & 3.7 & 66.0 & 8.99 & 20 \\ 
  HD 215806 & red & 45 & 1.30 & 0.17 & 1.53 & 71.5 & 4.3 & 67.0 & $-$ & 20 \\ 
  HILT 960 & blue & 80 & 4.17 & 0.57 & 5.69 & 56.0 & 4.2 & 54.9 & 10.96 & 100 \\ 
  HILT 960 & green & 80 & 3.78 & 0.57 & 5.21 & 55.5 & 3.7 & 54.5 & 9.84 & 100 \\ 
  HILT 960 & red & 80 & 3.39 & 0.42 & 4.46 & 55.0 & 3.7 & 54.0 & 9.49 & 100 \\ 
  VI CYG 12 & blue & 107 & 7.60 & 0.48 & 9.31 & 117.7 & 1.9 & 117.0 & 13.12 & 20 \\ 
  VI CYG 12 & green & 106 & 6.61 & 0.49 & 7.89 & 118.1 & 1.9 & 116.2 & 11.96 & 20 \\ 
  VI CYG 12 & red & 105 & 5.24 & 0.38 & 7.06 & 117.8 & 1.7 & 117.0 & 8.41 & 20 \\ 
   \hline
\end{tabular}
\end{table*}

\begin{knitrout}
\definecolor{shadecolor}{rgb}{0.969, 0.969, 0.969}\color{fgcolor}\begin{figure}
\includegraphics[width=\maxwidth]{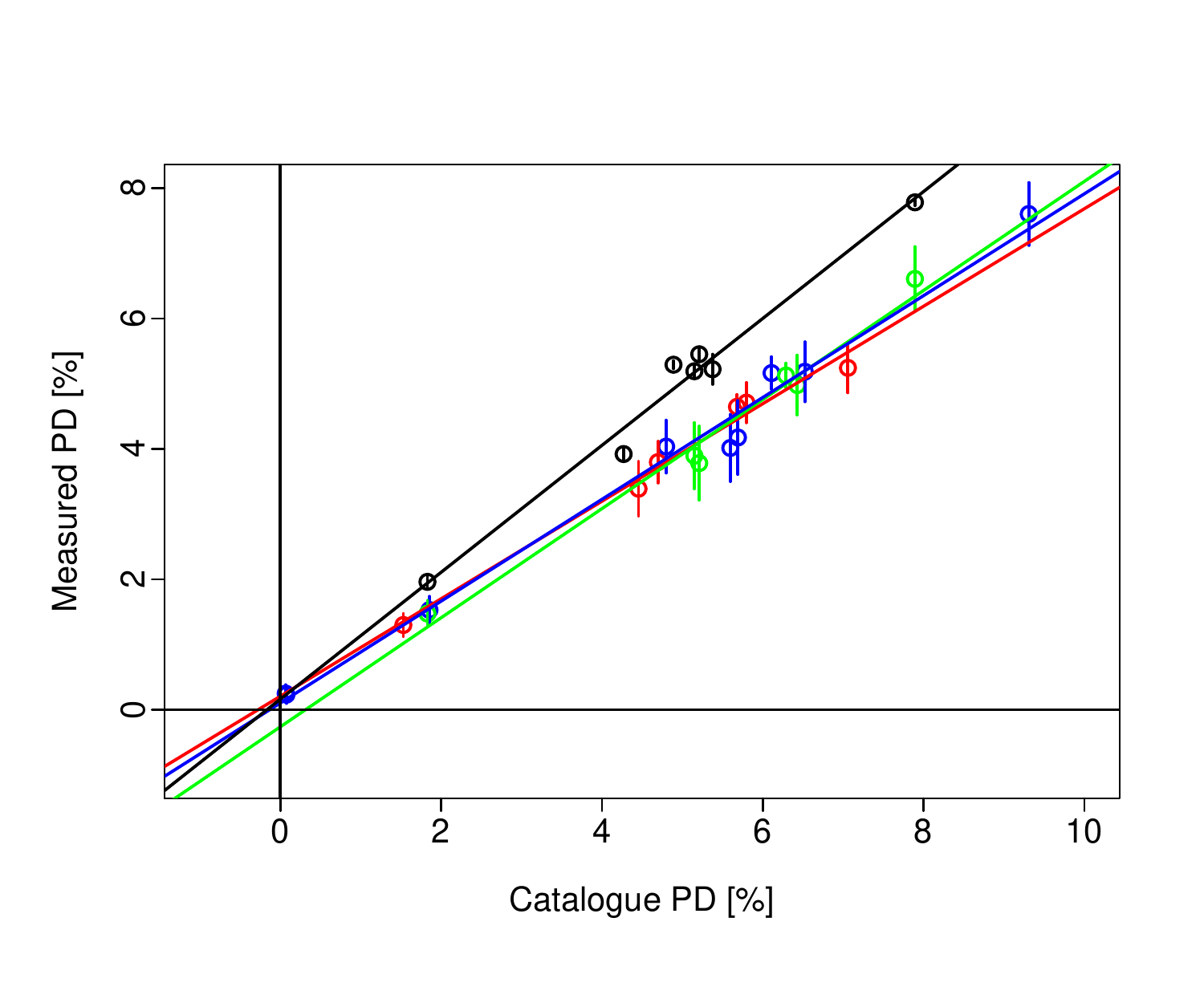} \caption[Measured PD as a function of catalogued PD for the blue, green and red cameras of the RINGO3 polarimeter (colour coded)]{Measured PD as a function of catalogued PD for the blue, green and red cameras of the RINGO3 polarimeter (colour coded). The colours of the RINGO3 blue, green and red cameras approximately correspond to the B+V, R and I, respectively. For comparison the RoboPol polarimeter data are also shown (black points and black solid line, after Table~1 of \cite{King2014}). RoboPol polarimeter works in the Johnson R-band.}\label{fig:lt_vs_robopol}
\end{figure}

\end{knitrout}

We have performed a long-term analysis of four polarimetric standard 
stars in order to calibrate the RINGO3 polarimeter operating at the 
Liverpool Telescope. Due to hardware changes, we divided the 
observations into five intervals. We have shown that the fifth interval 
provides the best calibrated results. Therefore we added five more
standard stars to our analysis that were recently added and observed
during the last epoch. Their measured polarisation degree as well as position angle
are shown in Fig.~\ref{fig:pd9} and Fig.~\ref{fig:pa9}, respectively.
Resulting values as well as the catalogue values are gathered in Table~\ref{tab:factors5}. 
For BD~+59~389 as well as for HD~14069 the gain was set to 100 before 57200 MJD and 
was changed to 20 after this date. Because both stars are brighter than 
10th magnitude the read noise is not really important.
Therefore it was basically the realuminization that improved the results
in the last epoch and the EMGAIN setting did not have such a big influence.

The relationship between the RINGO3 values and the catalogue ones are shown
in Fig.~\ref{fig:lt_vs_robopol}. For comparison the measured values of seven
high polarisation standards \citep[for details check the Table~1 of][]{King2014}
obtained with the Robopol polarimeter are shown. The relations between
the RINGO3 measured values and the catalogued ones are summarised as follows:
$$\mathrm{blue~(B+V): }~\mathrm{PD} = 0.7808 \cdot \mathrm{PD}_\mathrm{cat} + 0.1025,$$
$$\mathrm{green~(R): }~\mathrm{PD} = 0.8364 \cdot \mathrm{PD}_\mathrm{cat} \ensuremath{-0.263},$$
$$\mathrm{red~(I): }~\mathrm{PD} = 0.7475 \cdot \mathrm{PD}_\mathrm{cat} + 0.2061.$$

Table~\ref{tab:factors5} gives the average and standard deviation of nine standard stars 
(seven polarisation standards and two zero-polarised standards) measured 
during the fifth epoch. Because the standard deviation measures the 
scattering around a mean value, the results from Table~\ref{tab:factors5} can be used to 
check their stability with time.

Of the seven polarisation standard stars, BD~+25~727 is the one showing 
the most stable results, with the lowest value of standard deviation in 
the polarisation degree with $\sim 0.2$\%, while HILT~960 displays the 
highest dispersion with 0.57\%. 
Both zero-polarised standard stars display a rather constant polarisation degree over time. 
Although BD~+59~389 appears to be less variable than BD~+64~106, the former shows a 
correlation with the seeing conditions, which most likely results from the presence of a nearby star 
(Fig.~\ref{fig:cor_seeing}). BD~+64~106 appears to be more variable during the fifth 
epoch than it was before (see Fig.~\ref{fig:bd64_pd} and \ref{fig:bd64_pa}). More observations are 
needed to reveal if this star can be considered as a proper standard 
star. With respect to the polarisation angle, the scattering around the 
mean is $\lesssim 4^{\circ}$ for all sources.

The present work will be useful not only for RINGO3 users, but also as a 
reference for anyone performing polarimetric observations. With 
forthcoming big telescopes we are dramatically
missing the faint polarimetric standard stars. Therefore, we need to not only 
verify which of known standard stars are stable ones, but there is also an 
urgent need to look for new and fainter ones. As for RINGO3 in 
particular, the observers would greatly benefit if the list of standard 
stars is extended to cover a larger range in polarisation degree, especially for polarisation degree 
between 2\% and 4\%, as can be seen in Fig.~\ref{fig:lt_vs_robopol}.

The results of our study should be applied to all polarimetric observations performed 
with RINGO3 between December 7th, 2012 --- when the instrument was commissioned,
and January 7th, 2016 --- when the latest data presented in this publication were obtained. 
These dates correspond to 56268 MJD
and 57394 MJD, respectively.
The data presented in this publication will be available as a catalogue at 
VizieR\footnote{\url{http://vizier.u-strasbg.fr/vizier/}} service, that will be updated 
periodically as the polarimetric standard stars are observed and new data are available.

\section*{Acknowledgements}

We are grateful to the anonymous referee whose suggestions helped us to improve our paper significantly.
We are also very grateful to Frans Snik for his valuable comments and discussion as well as to Herv\'e Lamy 
for chairing the COST Action MP1104 and all the support and possibilities of extending our knowledge on polarisation that we got during this action.
We are grateful to Helen Jermak for technical assistance. 
This work has been supported by Polish National Science Centre grant
DEC-2011/03/D/ST9/00656 (AS, KK, M\.Z). This research was partly supported by 
the EU COST Action MP1104 "Polarization as a tool to study the solar system and
beyond" within STSM projects: COST-STSM-MP1104-14064, COST-STSM-MP1104-16823
COST-STSM-MP1104-14070 and COST-STSM-MP1104-16821. Data analysis and figures 
were partly prepared using R \citep{rcite}.
The Liverpool Telescope is operated on the island of La Palma by Liverpool John
Moores University in the Spanish Observatorio del Roque de los Muchachos of 
the Instituto de Astrofisica de Canarias with financial support from
the UK Science and Technology Facilities Council. 

\bibliographystyle{mn2e}
\bibliography{slowikowska}

\begin{thebibliography}{}

\bibitem[\protect\citeauthoryear{{Abdo}, {Ackermann}, {Ajello}, {Axelsson},
  {Baldini}, {Ballet}, {Barbiellini}, {Bastieri}, {Baughman}, {Bechtol} \& et
  al.}{{Abdo} et~al.}{2010}]{Abdo2010}
{Abdo} A.~A.,  {Ackermann} M.,  {Ajello} M.,  {Axelsson} M.,  {Baldini} L.,
  {Ballet} J.,  {Barbiellini} G.,  {Bastieri} D.,  {Baughman} B.~M.,  {Bechtol}
  K.,    et al. 2010, \nat, 463, 919

\bibitem[\protect\citeauthoryear{{Arnold}, {Steele}, {Bates}, {Mottram} \&
  {Smith}}{{Arnold} et~al.}{2012}]{Arnold2012}
{Arnold} D.~M.,  {Steele} I.~A.,  {Bates} S.~D.,  {Mottram} C.~J.,    {Smith}
  R.~J.,  2012, in Society of Photo-Optical Instrumentation Engineers (SPIE)
  Conference Series Vol.~8446 of Society of Photo-Optical Instrumentation
  Engineers (SPIE) Conference Series, {RINGO3: a multi-colour fast response
  polarimeter}.
p.~2

\bibitem[\protect\citeauthoryear{{Bertin} \& {Arnouts}}{{Bertin} \&
  {Arnouts}}{1996}]{Bertin1996}
{Bertin} E.,  {Arnouts} S.,  1996, \aaps, 117, 393

\bibitem[\protect\citeauthoryear{{Graham}, {Kalas} \& {Matthews}}{{Graham}
  et~al.}{2007}]{Graham2007}
{Graham} J.~R.,  {Kalas} P.~G.,    {Matthews} B.~C.,  2007, \apj, 654, 595

\bibitem[\protect\citeauthoryear{{Hansen} \& {Hovenier}}{{Hansen} \&
  {Hovenier}}{1974}]{Hansen1974}
{Hansen} J.~E.,  {Hovenier} J.~W.,  1974, Journal of Atmospheric Sciences, 31,
  1137

\bibitem[\protect\citeauthoryear{{King}, {Blinov}, {Ramaprakash}, {Myserlis} \&
  et al.}{{King} et~al.}{2014}]{King2014}
{King} O.~G.,  {Blinov} D.,  {Ramaprakash} A.~N.,  {Myserlis} I.,    et al. A.,
   2014, \mnras, 442, 1706

\bibitem[\protect\citeauthoryear{{Moran}, {Shearer}, {Mignani},
  {S{\l}owikowska}, {De Luca}, {Gouiff{\`e}s} \& {Laurent}}{{Moran}
  et~al.}{2013}]{Moran2013}
{Moran} P.,  {Shearer} A.,  {Mignani} R.~P.,  {S{\l}owikowska} A.,  {De Luca}
  A.,  {Gouiff{\`e}s} C.,    {Laurent} P.,  2013, \mnras, 433, 2564

\bibitem[\protect\citeauthoryear{{Mundell}, {Kopa{\v c}}, {Arnold}, {Steele},
  {Gomboc}, {Kobayashi}, {Harrison}, {Smith}, {Guidorzi}, {Virgili}, {Melandri}
  \& {Japelj}}{{Mundell} et~al.}{2013}]{Mundell2013}
{Mundell} C.~G.,  {Kopa{\v c}} D.,  {Arnold} D.~M.,  {Steele} I.~A.,  {Gomboc}
  A.,  {Kobayashi} S.,  {Harrison} R.~M.,  {Smith} R.~J.,  {Guidorzi} C.,
  {Virgili} F.~J.,  {Melandri} A.,    {Japelj} J.,  2013, \nat, 504, 119

\bibitem[\protect\citeauthoryear{{R Core Team}}{{R Core Team}}{2013}]{rcite}
{R Core Team} 2013, R: A Language and Environment for Statistical Computing.
R Foundation for Statistical Computing, Vienna, Austria

\bibitem[\protect\citeauthoryear{{Schmidt}, {Elston} \& {Lupie}}{{Schmidt}
  et~al.}{1992}]{Schmidt1992}
{Schmidt} G.~D.,  {Elston} R.,    {Lupie} O.~L.,  1992, \aj, 104, 1563

\bibitem[\protect\citeauthoryear{{S{\l}owikowska}, {Kanbach}, {Kramer} \&
  {Stefanescu}}{{S{\l}owikowska} et~al.}{2009}]{Slowikowska2009}
{S{\l}owikowska} A.,  {Kanbach} G.,  {Kramer} M.,    {Stefanescu} A.,  2009,
  \mnras, 397, 103

\bibitem[\protect\citeauthoryear{{Sparks} \& {Axon}}{{Sparks} \&
  {Axon}}{1999}]{Sparks1999}
{Sparks} W.~B.,  {Axon} D.~J.,  1999, \pasp, 111, 1298

\bibitem[\protect\citeauthoryear{{Steele}, {Smith}, {Rees},  et~al.,}{{Steele}
  et~al.}{2004}]{Steele2004}
{Steele} I.~A.,  {Smith} R.~J.,  {Rees} P.~C.,     et~al., 2004, in {Oschmann}
  Jr. J.~M.,  ed., Ground-based Telescopes Vol.~5489 of Society of
  Photo-Optical Instrumentation Engineers (SPIE) Conference Series, {The
  Liverpool Telescope: performance and first results}.
pp 679--692

\bibitem[\protect\citeauthoryear{{Turnshek}, {Bohlin}, {Williamson} II,
  {Lupie}, {Koornneef} \& {Morgan}}{{Turnshek} et~al.}{1990}]{Turnshek1990}
{Turnshek} D.~A.,  {Bohlin} R.~C.,  {Williamson} II R.~L.,  {Lupie} O.~L.,
  {Koornneef} J.,    {Morgan} D.~H.,  1990, \aj, 99, 1243

\bibitem[\protect\citeauthoryear{{Whittet}, {Martin}, {Hough}, {Rouse},
  {Bailey} \& {Axon}}{{Whittet} et~al.}{1992}]{Whittet1992}
{Whittet} D.~C.~B.,  {Martin} P.~G.,  {Hough} J.~H.,  {Rouse} M.~F.,  {Bailey}
  J.~A.,    {Axon} D.~J.,  1992, \apj, 386, 562

\end{thebibliography}

\appendix
\section{Appendix}
\label{sec:appendix}

\subsection{Polarisation calculations and error propagation}

\subsubsection{Sparks \& Axon}
Stokes from Sparks \& Axon: $I_i$, $Q_i$, $U_i$, $\sigma_{I_i}$, $\sigma_{Q_i}$ and $\sigma_{U_i}$, 
where $i$ denotes single observation, i.e. intensity measured in 8 polariser positions. Let us define 
$q_i = Q_i / I_i$ and $u_i = U_i / I_i$ with respective errors 
$\sigma_{q_i} = \sqrt{(\frac{1}{I_i}\sigma_{Q_i})^2 + (-\frac{Q_i}{I^2_i} \sigma_{I_i})^2}$
and $\sigma_{u_i} = \sqrt{(\frac{1}{I_i}\sigma_{U_i})^2 + (-\frac{U_i}{I^2_i} \sigma_{I_i})^2}$.\\

\subsubsection{Shifts}
For non--polarised standard stars $q^s_i = q_i - \overline{q}$, $u^s_i = u_i - \overline{u}$ where 
means are weighted means calculated from observations from the same MJD range between LT hardware changes:
$$\overline{q} = \frac{\sum_i q_i/\sigma_{q_i}^2}   {\sum_i 1/\sigma_{q_i}^2},
\overline{u} = \frac{\sum_i u_i /\sigma_{u_i}^2}   {\sum_i 1/\sigma_{u_i}^2}$$

$$\sigma_{\overline{q}} = \sqrt{\frac{1}{\sum_i 1 / \sigma_{q_i}^{2}}}, 
\sigma_{\overline{u}} = \sqrt{\frac{1}{\sum_i 1/\sigma_{u_i}^{2}}}$$
Shifted $q^s_i$ and $u^s_i$ with their corresponding errors:
$$q^s_i = q_i - \overline{q}, u^s_i = u_i - \overline{u}$$
$$\sigma_{q^s_i} = \sqrt{\sigma_{q_i}^2 + \sigma_{\overline{q}}^2}, \sigma_{u^s_i} = \sqrt{\sigma_{u_i}^2 + \sigma_{\overline{u}}^2}$$

Instrumental polarisation:
$$\mathrm{IP} = \sqrt{\overline{q}^2 + \overline{u}^2}$$
$$\sigma_\mathrm{IP} = \sqrt{\frac{\overline{q}^2}{\overline{q}^2 + \overline{u}^2}\sigma_{\overline{q}}^2 + 
\frac{\overline{u}^2}{\overline{q}^2 + \overline{u}^2}\sigma_{\overline{u}}^2}$$

Polarisation degree (PD$_i$) calculated from $q^s_i$ and $u^s_i$:
$$\mathrm{PD}_i = \sqrt{(q^s_i)^2 + (u^s_i)^2} \cdot 100 \%$$
$$\sigma_{\mathrm{PD}_i} = \sqrt{\frac{(q^s_i)^2}{(q^s_i)^2 + (u^s_i)^2} \sigma^2_{q^s_i} + 
\frac{(u^s_i)^2}{(q^s_i)^2 + (u^s_i)^2} \sigma^2_{u^s_i}} \cdot 100 \% $$

Polarisation angle (PA$_i$) calculated from $q^s_i$ and $u^s_i$:
$$\mathrm{PA}_i = \frac{1}{2} \arctan \left( \frac{u^s_i}{q^s_i} \right) \cdot \frac{180^\circ}{\pi}$$
$$\sigma_{\mathrm{PA}_i} = \sqrt{\left(\frac{1}{2q^s_i (1 + (u^s_i)^2 / (q^s_i)^2)}\right)^2 \sigma^2_{u^s_i} + 
\left( -\frac{u^s_i}{2((q^s_i)^2 + (u^s_i)^2)} \right)^2 \sigma^2_{q^s_i}} \cdot \frac{180^\circ}{\pi}$$

\begin{table*}
 \scriptsize
 \caption{Linear fits of Q/I as a function of MJD in the form of $\mathrm{Q/I} = a \cdot \mathrm{MJD} + b$. There are not enough data points in case of HD~14069 in green colour in the first epoch.}
 \label{tab:qi_fits}
 \begin{tabular}{llrrrrrr}
   \hline
     \multirow{3}{*}{MJD range} & \multirow{3}{*}{Source} & \multicolumn{6}{c}{Q / I}\\
         & & \multicolumn{2}{c}{blue} & \multicolumn{2}{c}{green} & \multicolumn{2}{c}{red} \\
         & & \multicolumn{1}{c}{a [10$^{-5}$]} & \multicolumn{1}{c}{b} & \multicolumn{1}{c}{a [10$^{-5}$]} & \multicolumn{1}{c}{b} & \multicolumn{1}{c}{a [10$^{-5}$]} & \multicolumn{1}{c}{b} \\
     \hline
56200 - 56315 &G~191-B2B
 &$-6.22588 \pm 3.7067$ &$3.4655 \pm 2.08656$ &$28.22669 \pm 18.88837$ &$-15.86979 \pm 10.62995$ &$28.26401 \pm 11.61088$ &$-15.82304 \pm 6.53599$\\
 &HD~14069
 &$-17.02482 \pm 15.08858$ &$9.54221 \pm 8.49333$ &--- &--- &$12.02995 \pm 9.76139$ &$-6.71683 \pm 5.49481$\\
56315 - 56638 &G~191-B2B
 &$0.24016 \pm 0.19987$ &$-0.17684 \pm 0.11299$ &$0.76741 \pm 0.24999$ &$-0.40406 \pm 0.14132$ &$-0.04385 \pm 0.52521$ &$0.12577 \pm 0.29691$\\
 &HD~14069
 &$1.04026 \pm 0.49436$ &$-0.62949 \pm 0.27956$ &$0.62723 \pm 0.31245$ &$-0.34174 \pm 0.17669$ &$0.08544 \pm 0.46375$ &$0.0132 \pm 0.26224$\\
56638 - 56816 &G~191-B2B
 &$0.72573 \pm 0.93509$ &$-0.42874 \pm 0.53014$ &$-1.16802 \pm 1.69196$ &$0.62793 \pm 0.95924$ &$-1.45806 \pm 2.20054$ &$0.78977 \pm 1.24758$\\
 &HD~14069
 &$3.82252 \pm 1.55795$ &$-2.18277 \pm 0.88297$ &$4.06072 \pm 1.70524$ &$-2.33404 \pm 0.96647$ &$4.34963 \pm 1.62387$ &$-2.49979 \pm 0.92038$\\
56816 - 57200 &G~191-B2B
 &$-0.39898 \pm 0.25109$ &$0.23392 \pm 0.143$ &$0.30698 \pm 0.38481$ &$-0.16762 \pm 0.21916$ &$-0.29263 \pm 0.54787$ &$0.17182 \pm 0.31202$\\
 &HD~14069
 &$-0.93964 \pm 0.34431$ &$0.54138 \pm 0.19601$ &$-0.63593 \pm 0.33158$ &$0.36966 \pm 0.18875$ &$-0.2452 \pm 0.34043$ &$0.1463 \pm 0.19379$\\
57200 - 57400 &G~191-B2B
 &$0.07742 \pm 0.40258$ &$-0.06033 \pm 0.23069$ &$0.6248 \pm 0.62988$ &$-0.38828 \pm 0.36095$ &$0.6673 \pm 0.89742$ &$-0.40973 \pm 0.51425$\\
 &HD~14069
 &$1.01692 \pm 0.4484$ &$-0.59958 \pm 0.2569$ &$0.78509 \pm 0.48007$ &$-0.48009 \pm 0.27504$ &$1.32394 \pm 0.45557$ &$-0.78934 \pm 0.261$\\
\hline
 \end{tabular}

 \end{table*}

\begin{table*}
 \scriptsize
 \caption{Linear fits of U/I as a function of MJD in the form of $\mathrm{U/I} = a \cdot \mathrm{MJD} + b$. There are not enough data points in case of HD~14069 in green colour in the first epoch.}
 \label{tab:ui_fits}
 \begin{tabular}{llrrrrrr}
   \hline
     \multirow{3}{*}{MJD range} & \multirow{3}{*}{Source} & \multicolumn{6}{c}{U / I}\\
         & & \multicolumn{2}{c}{blue} & \multicolumn{2}{c}{green} & \multicolumn{2}{c}{red} \\
         & & \multicolumn{1}{c}{a [10$^{-5}$]} & \multicolumn{1}{c}{b} & \multicolumn{1}{c}{a [10$^{-5}$]} & \multicolumn{1}{c}{b} & \multicolumn{1}{c}{a [10$^{-5}$]} & \multicolumn{1}{c}{b} \\
     \hline
56200 - 56315 &G~191-B2B
 &$9.74502 \pm 5.54679$ &$-5.50947 \pm 3.12238$ &$5.868 \pm 14.29402$ &$-3.28986 \pm 8.04435$ &$20.2827 \pm 11.30336$ &$-11.37728 \pm 6.36288$\\
 &HD~14069
 &$17.61409 \pm 7.36892$ &$-9.93589 \pm 4.14795$ &--- &--- &$21.55379 \pm 8.76519$ &$-12.10408 \pm 4.93404$\\
56315 - 56638 &G~191-B2B
 &$-1.80915 \pm 0.28135$ &$1.00956 \pm 0.15905$ &$-0.55277 \pm 0.31311$ &$0.32767 \pm 0.177$ &$2.58749 \pm 0.39251$ &$-1.4168 \pm 0.22189$\\
 &HD~14069
 &$-0.61845 \pm 0.44185$ &$0.33511 \pm 0.24987$ &$-0.31433 \pm 0.32345$ &$0.1861 \pm 0.18291$ &$2.14438 \pm 0.28758$ &$-1.18256 \pm 0.16262$\\
56638 - 56816 &G~191-B2B
 &$0.66244 \pm 0.82917$ &$-0.37935 \pm 0.47009$ &$-2.64107 \pm 1.50799$ &$1.48532 \pm 0.85494$ &$0.38329 \pm 1.89424$ &$-0.23 \pm 1.07392$\\
 &HD~14069
 &$-2.55777 \pm 1.32826$ &$1.44693 \pm 0.75279$ &$-0.46136 \pm 1.10551$ &$0.25078 \pm 0.62656$ &$2.13545 \pm 1.34432$ &$-1.22301 \pm 0.76193$\\
56816 - 57200 &G~191-B2B
 &$-0.09761 \pm 0.25007$ &$0.07728 \pm 0.14243$ &$0.13613 \pm 0.34261$ &$-0.04197 \pm 0.19513$ &$0.16111 \pm 0.55307$ &$-0.05785 \pm 0.31498$\\
 &HD~14069
 &$-1.90078 \pm 0.41329$ &$1.10415 \pm 0.23527$ &$-0.9641 \pm 0.38435$ &$0.58538 \pm 0.21879$ &$-1.1615 \pm 0.44841$ &$0.69734 \pm 0.25526$\\
57200 - 57400 &G~191-B2B
 &$-0.36752 \pm 0.4402$ &$0.22032 \pm 0.25225$ &$-0.23515 \pm 0.61656$ &$0.14996 \pm 0.35331$ &$-0.44733 \pm 1.09518$ &$0.26849 \pm 0.62757$\\
 &HD~14069
 &$-0.42762 \pm 0.47583$ &$0.25508 \pm 0.27262$ &$-0.02817 \pm 0.62404$ &$0.03215 \pm 0.35752$ &$-0.34508 \pm 0.59687$ &$0.21232 \pm 0.34196$\\
\hline
 \end{tabular}

 \end{table*}

\newpage

\begin{knitrout}
\definecolor{shadecolor}{rgb}{0.969, 0.969, 0.969}\color{fgcolor}\begin{figure}
\includegraphics[width=\maxwidth]{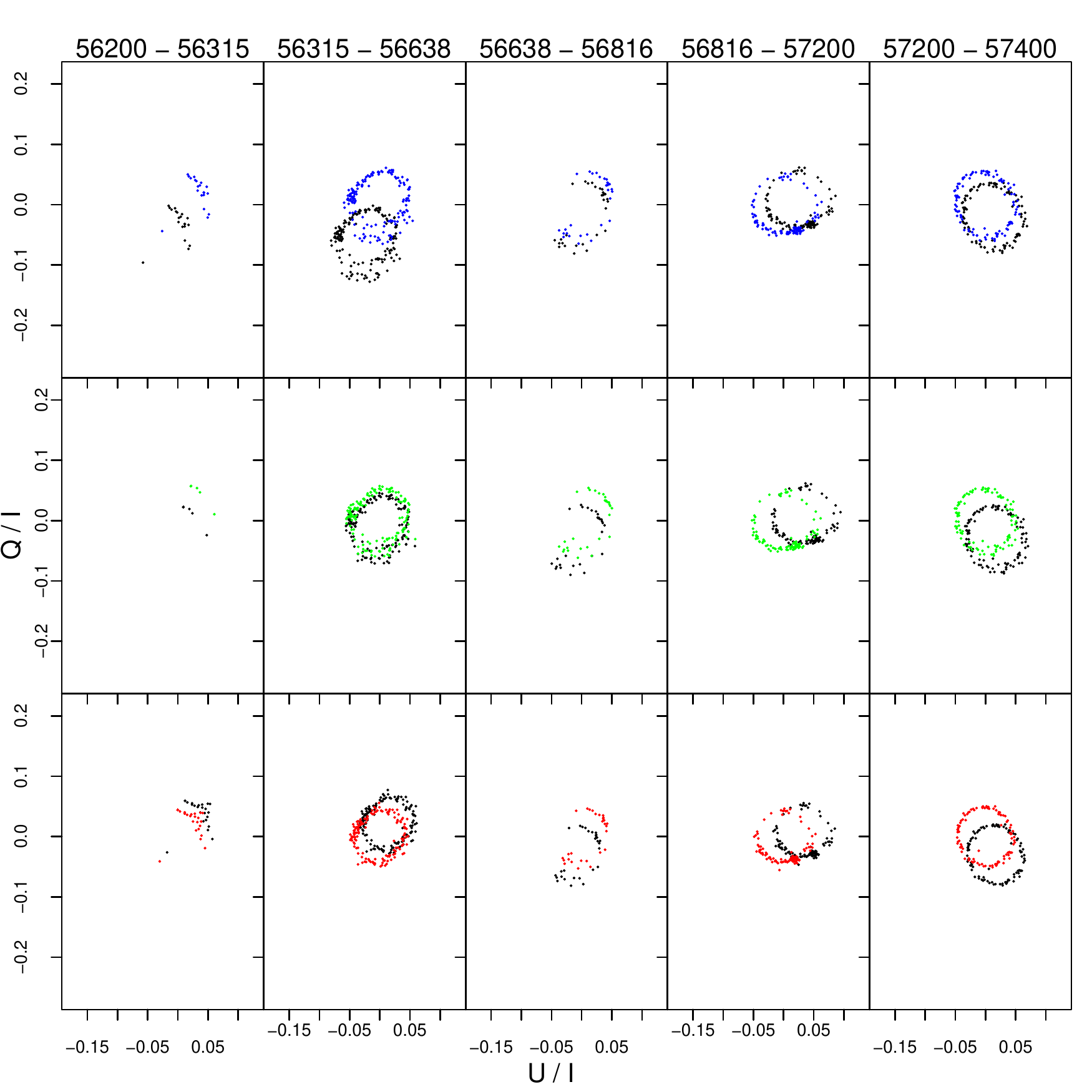} \caption[Normalised Q/I, U/I diagram of BD~+59~389 for five time span ranges for the blue, green and red cameras from the top to the bottom, respectively]{Normalised Q/I, U/I diagram of BD~+59~389 for five time span ranges for the blue, green and red cameras from the top to the bottom, respectively. Black points denote measurements before shifts while colour ones are shifted to (0, 0) origin and correspond to the respective camera. For polarised source the points follow a circle because of the Alt-Az mount of the telescope.}\label{fig:bd59_shift}
\end{figure}

\end{knitrout}

\begin{knitrout}
\definecolor{shadecolor}{rgb}{0.969, 0.969, 0.969}\color{fgcolor}\begin{figure}
\includegraphics[width=\maxwidth]{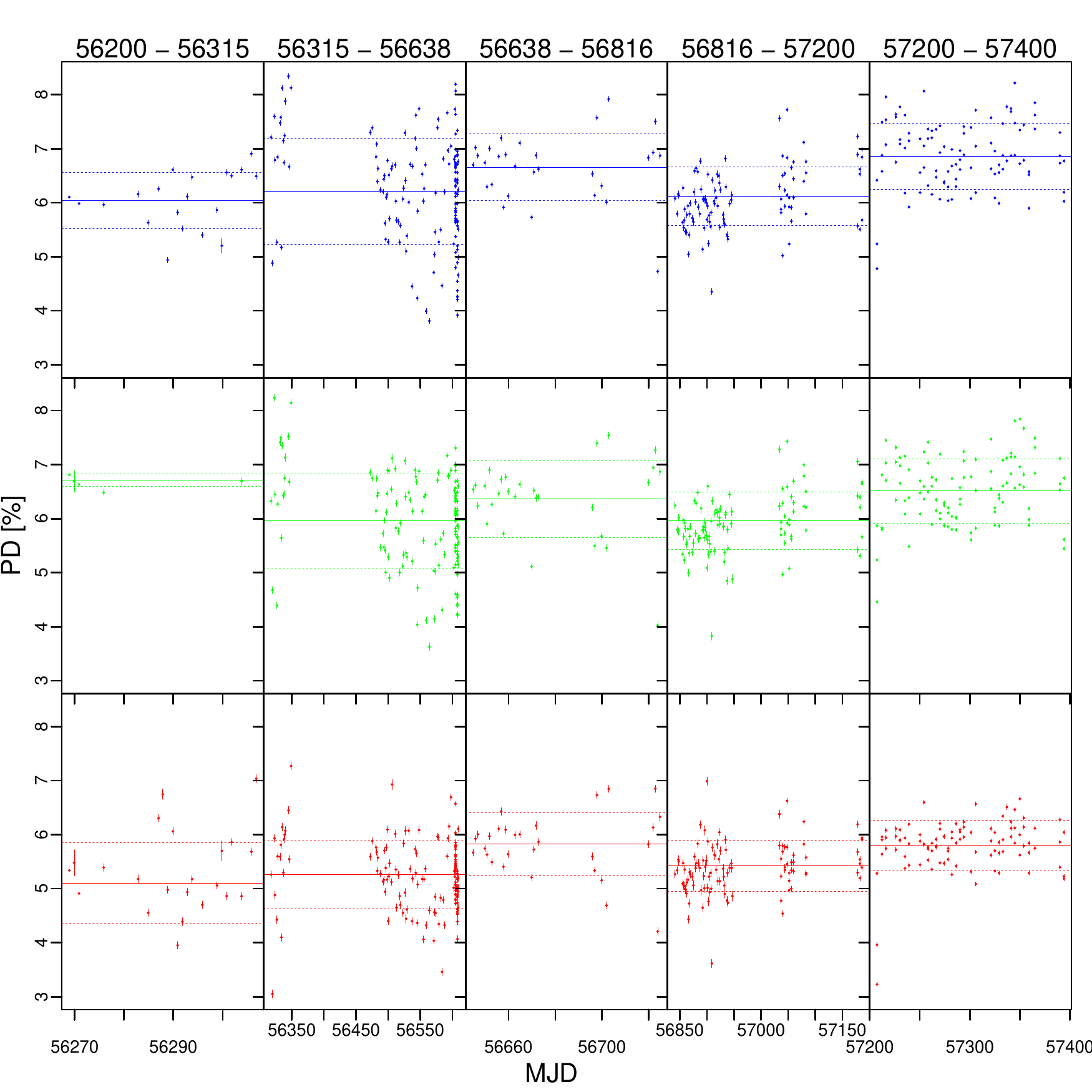} \caption[Resulting PD of BD~+59~389 for five time span ranges for the blue, green and red cameras from the top to the bottom, respectively]{Resulting PD of BD~+59~389 for five time span ranges for the blue, green and red cameras from the top to the bottom, respectively. Solid lines correspond to the mean values, while dotted lines to one standard deviation in each single panel. There are not many data points from the green camera in the first epoch, therefore the PD is not well constrained. }\label{fig:bd59_pd}
\end{figure}

\end{knitrout}

\begin{knitrout}
\definecolor{shadecolor}{rgb}{0.969, 0.969, 0.969}\color{fgcolor}\begin{figure}
\includegraphics[width=\maxwidth]{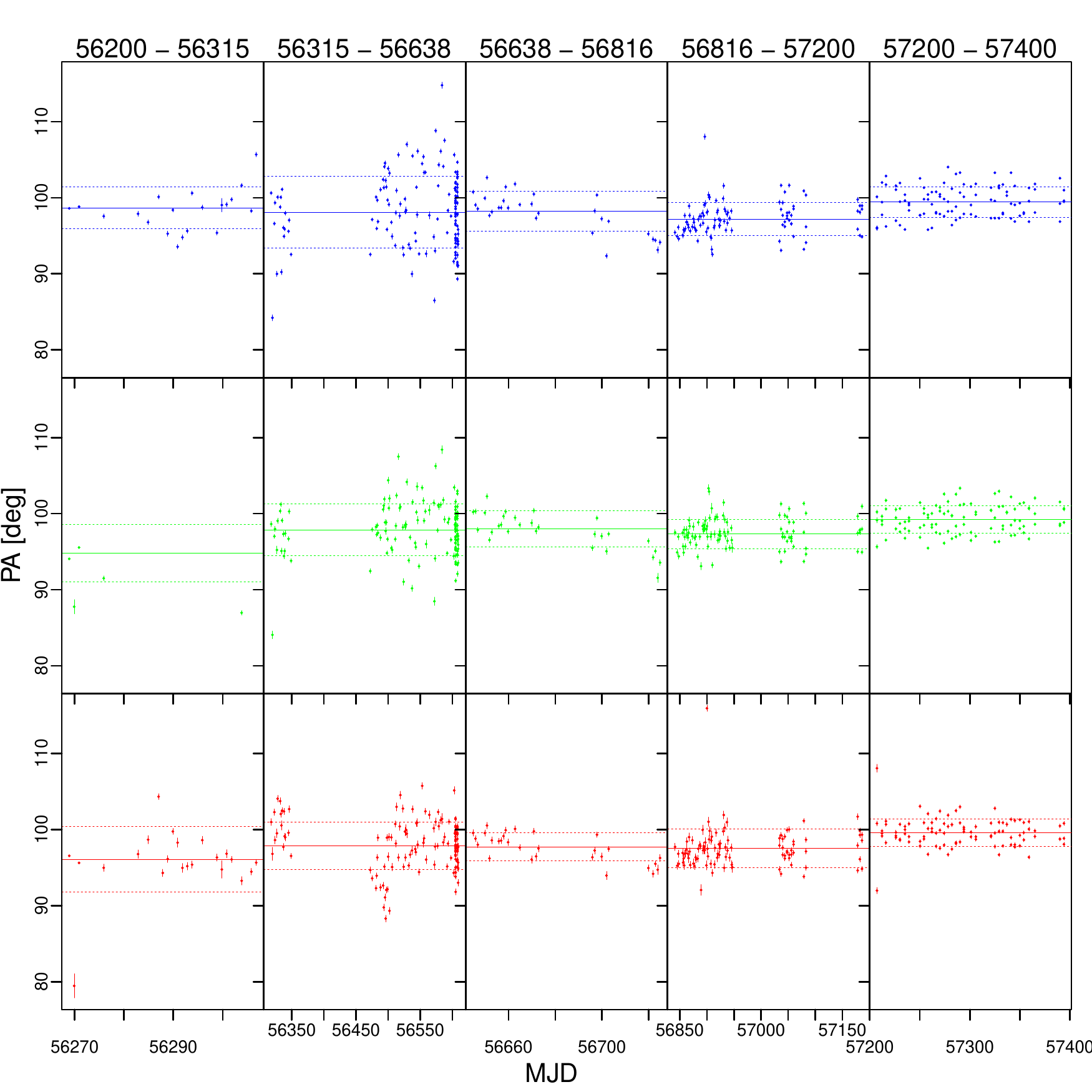} \caption[Resulting PA of BD~+59~389 for five time span ranges for the blue, green and red cameras from the top to the bottom, respectively]{Resulting PA of BD~+59~389 for five time span ranges for the blue, green and red cameras from the top to the bottom, respectively. Solid lines correspond to the mean values, while dotted lines to one standard deviation in each single panel. There are not many data points from the green camera in the first epoch, therefore the PA is not well constrained.}\label{fig:bd59_pa}
\end{figure}

\end{knitrout}

\begin{knitrout}
\definecolor{shadecolor}{rgb}{0.969, 0.969, 0.969}\color{fgcolor}\begin{figure}
\includegraphics[width=\maxwidth]{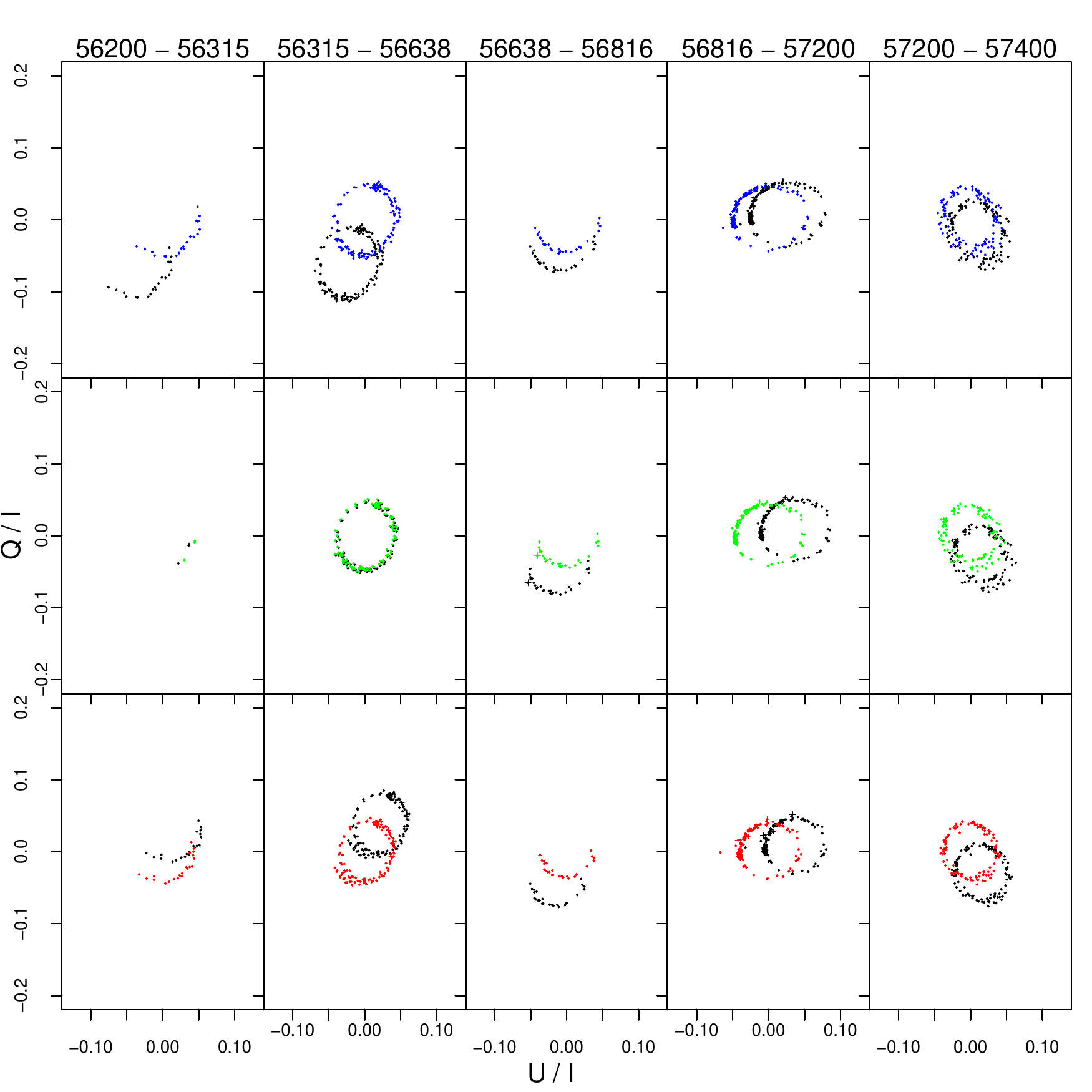} \caption[Normalised Q/I, U/I diagram of BD~+64~106  for five time span ranges for the blue, green and red cameras from the top to the bottom, respectively]{Normalised Q/I, U/I diagram of BD~+64~106  for five time span ranges for the blue, green and red cameras from the top to the bottom, respectively. Black points denote measurements before shifts while colour ones are shifted to (0, 0) origin and correspond to the respective camera. For polarised source the points follow a circle because of the Alt-Az mount of the telescope.}\label{fig:bd64_shift}
\end{figure}

\end{knitrout}

\begin{knitrout}
\definecolor{shadecolor}{rgb}{0.969, 0.969, 0.969}\color{fgcolor}\begin{figure}
\includegraphics[width=\maxwidth]{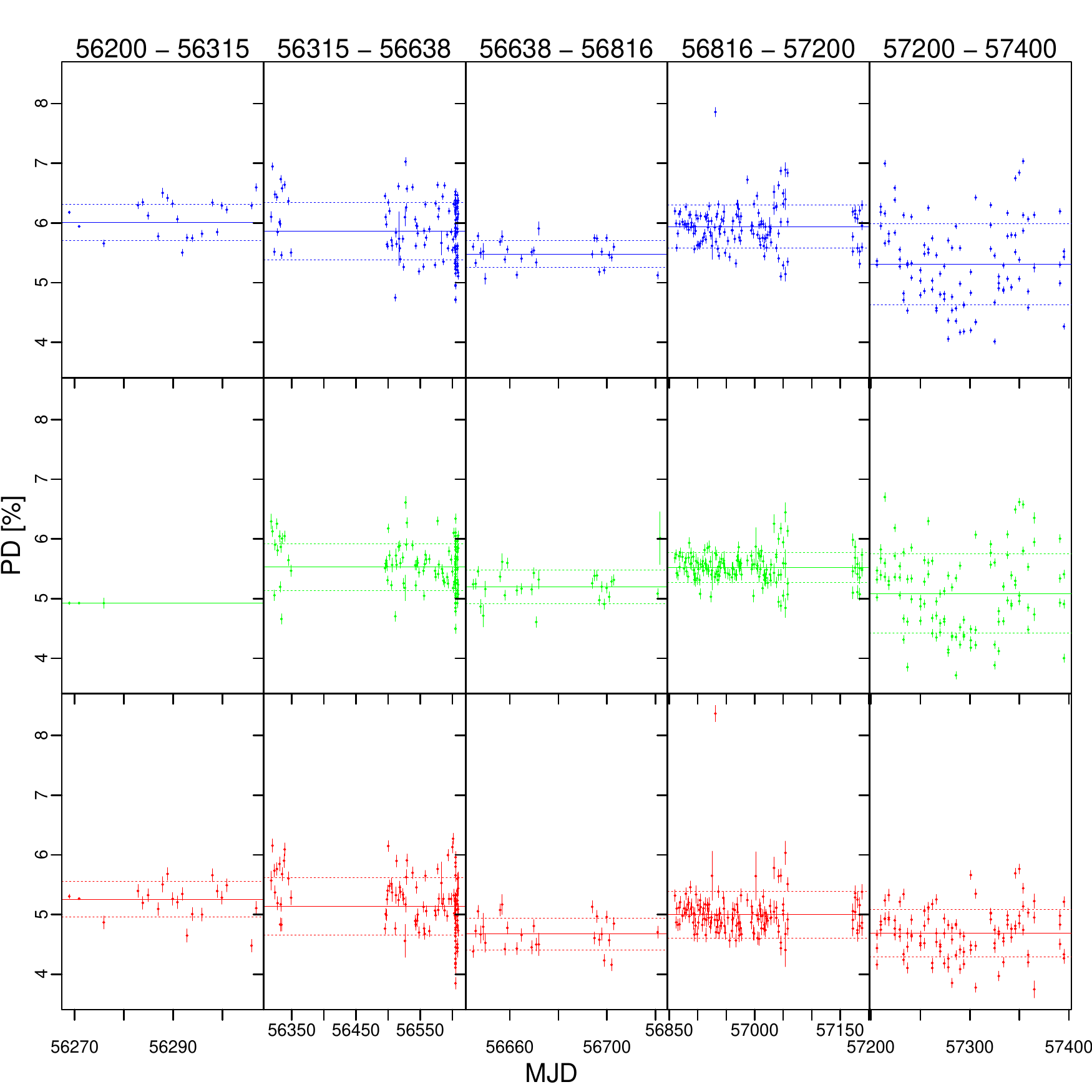} \caption[Resulting PD of BD~+64~106 for five time span ranges  for five time span ranges for the blue, green and red cameras from the top to the bottom, respectively]{Resulting PD of BD~+64~106 for five time span ranges  for five time span ranges for the blue, green and red cameras from the top to the bottom, respectively. Solid lines correspond to the mean values, while dotted lines to one standard deviation in each single panel. There are not many data points from the green camera in the first epoch, therefore the PD is not well constrained.}\label{fig:bd64_pd}
\end{figure}

\end{knitrout}

\begin{knitrout}
\definecolor{shadecolor}{rgb}{0.969, 0.969, 0.969}\color{fgcolor}\begin{figure}
\includegraphics[width=\maxwidth]{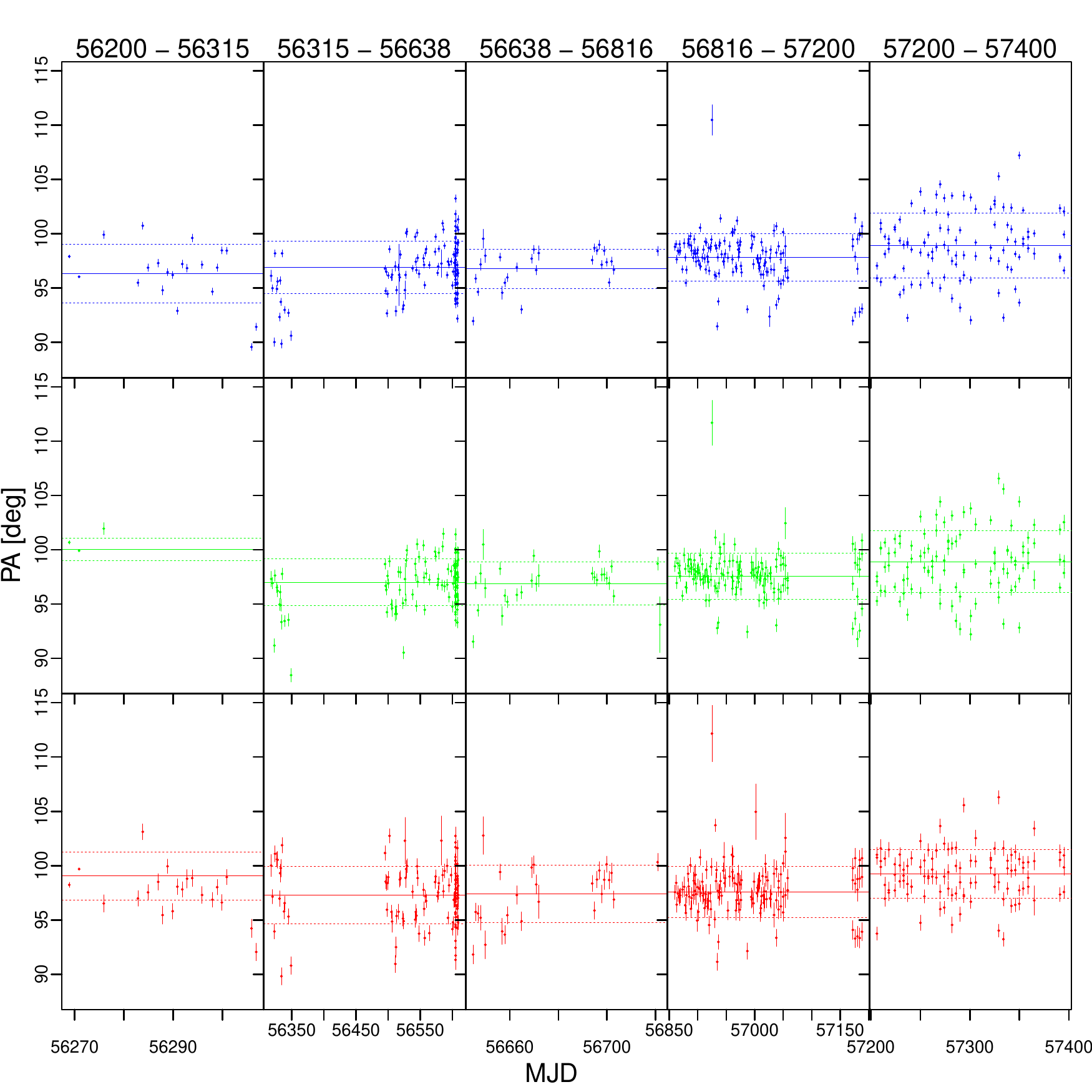} \caption[Resulting PA of BD~+64~106 for five time span ranges for the blue, green and red cameras from the top to the bottom, respectively]{Resulting PA of BD~+64~106 for five time span ranges for the blue, green and red cameras from the top to the bottom, respectively. Solid lines correspond to the mean values, while dotted lines to one standard deviation in each single panel. There are not many data points from the green camera in the first epoch, therefore the PA is not well constrained.}\label{fig:bd64_pa}
\end{figure}

\end{knitrout}

\begin{knitrout}
\definecolor{shadecolor}{rgb}{0.969, 0.969, 0.969}\color{fgcolor}\begin{figure}
\includegraphics[width=\maxwidth]{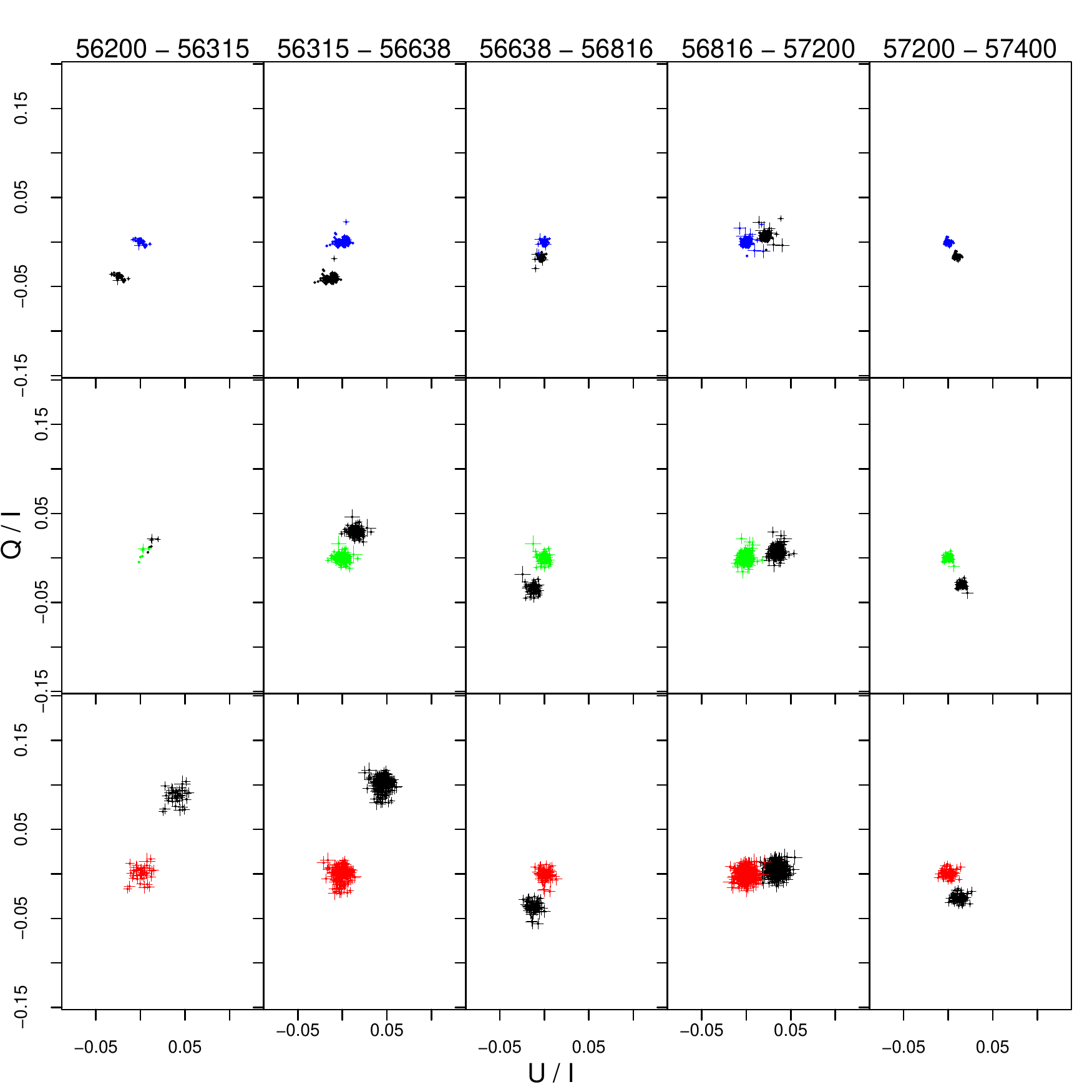} \caption[Normalised Q/I, U/I diagram of G~191-B2B for five time span ranges for the blue, green and red cameras from the top to the bottom, respectively]{Normalised Q/I, U/I diagram of G~191-B2B for five time span ranges for the blue, green and red cameras from the top to the bottom, respectively. Black points denote measurements before shifts while colour ones are shifted to (0, 0) origin and correspond to the respective camera.}\label{fig:g191_shift}
\end{figure}

\end{knitrout}

\begin{knitrout}
\definecolor{shadecolor}{rgb}{0.969, 0.969, 0.969}\color{fgcolor}\begin{figure}
\includegraphics[width=\maxwidth]{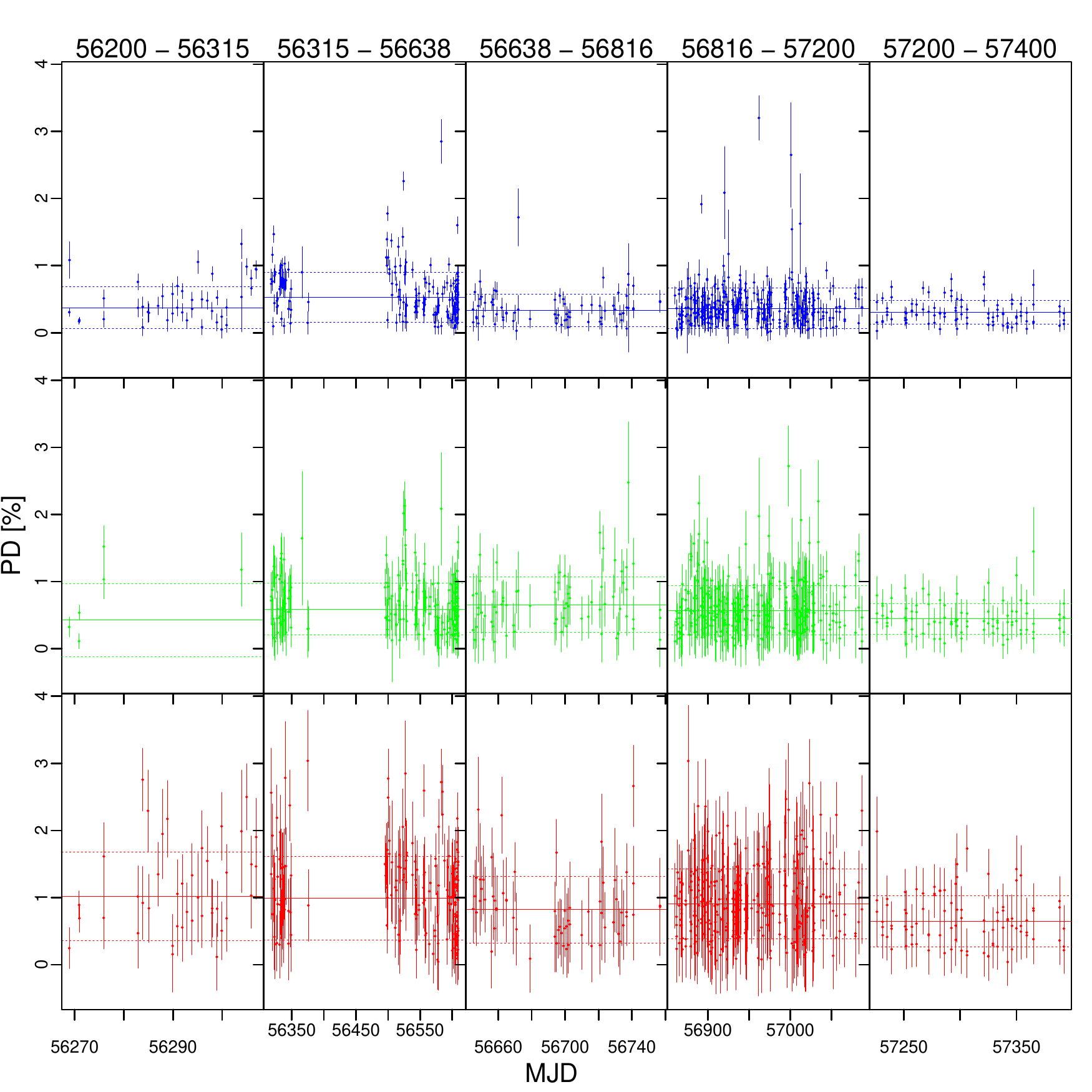} \caption[Resulting PD of G~191-B2B for five time span ranges for the blue, green and red cameras from the top to the bottom, respectively]{Resulting PD of G~191-B2B for five time span ranges for the blue, green and red cameras from the top to the bottom, respectively. Solid lines correspond to the mean values, while dotted lines to one standard deviation in each single panel.}\label{fig:g191_pd}
\end{figure}

\end{knitrout}

\begin{knitrout}
\definecolor{shadecolor}{rgb}{0.969, 0.969, 0.969}\color{fgcolor}\begin{figure}
\includegraphics[width=\maxwidth]{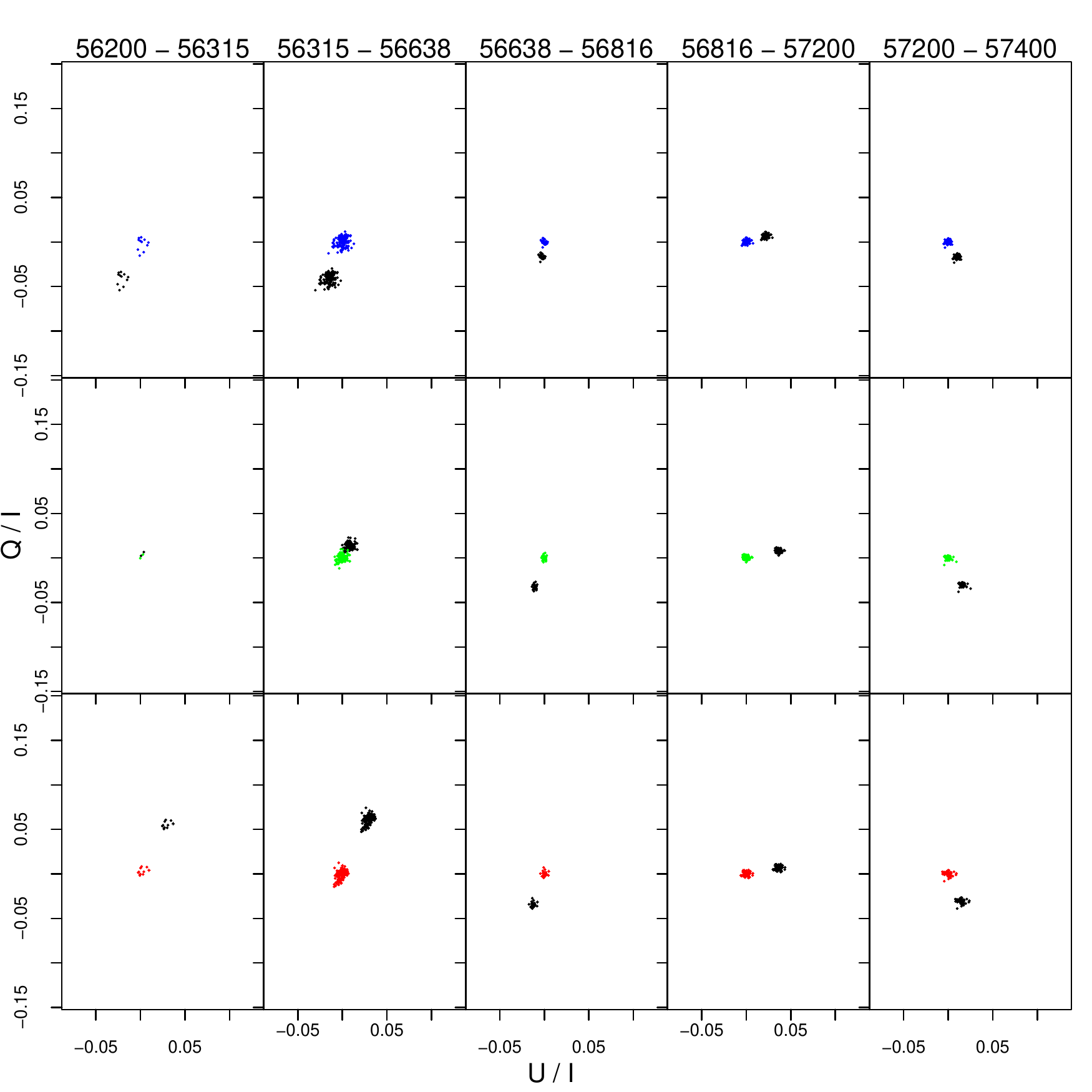} \caption[Normalised Q/I, U/I diagram of HD~14069 for five time span ranges for five time span ranges for the blue, green and red cameras from the top to the bottom, respectively]{Normalised Q/I, U/I diagram of HD~14069 for five time span ranges for five time span ranges for the blue, green and red cameras from the top to the bottom, respectively. Black points denote measurements before shifts while colour ones are shifted to (0, 0) origin and correspond to the respective camera.}\label{fig:hd14_shift}
\end{figure}

\end{knitrout}

\begin{knitrout}
\definecolor{shadecolor}{rgb}{0.969, 0.969, 0.969}\color{fgcolor}\begin{figure}
\includegraphics[width=\maxwidth]{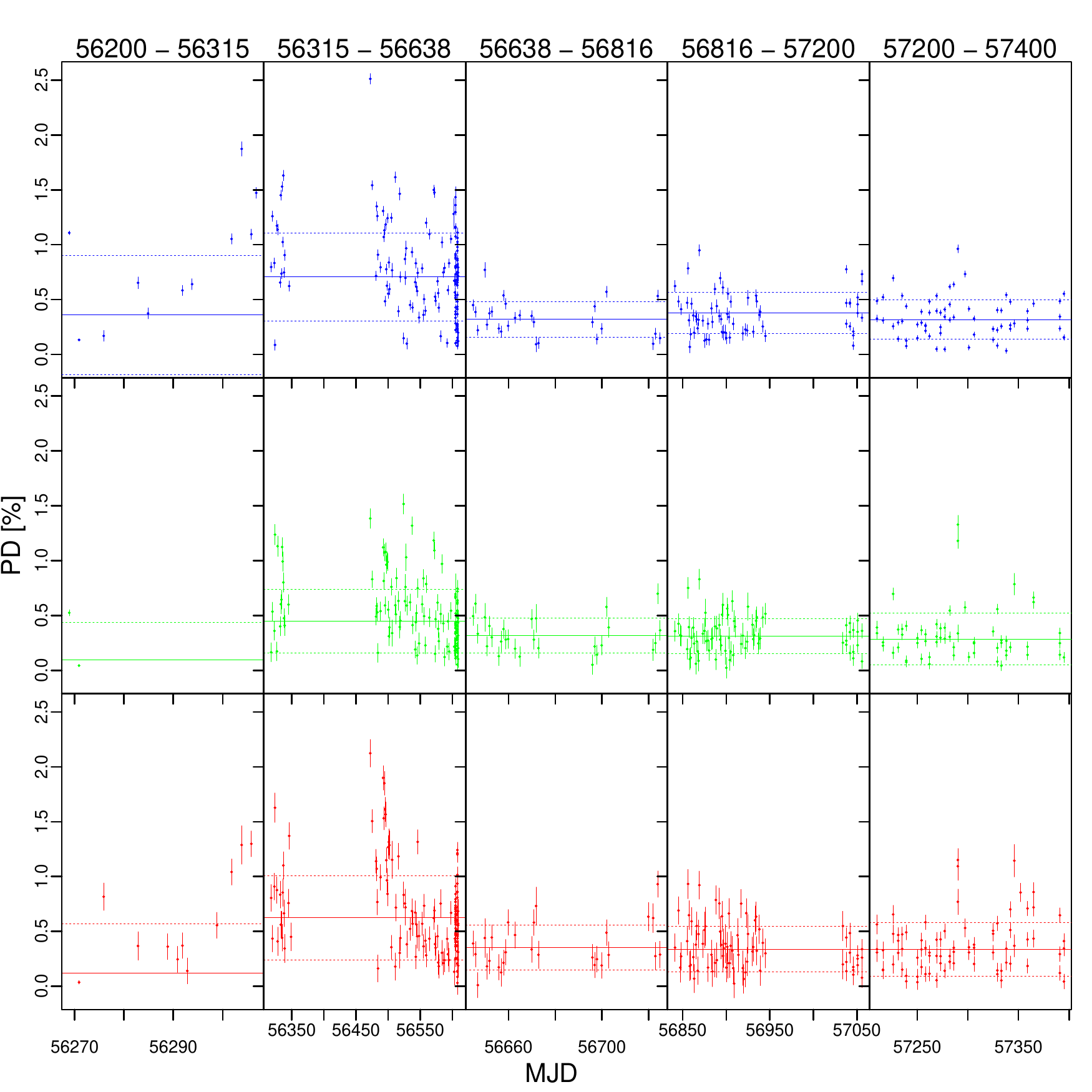} \caption[Resulting PD of HD~14069 for five time span ranges for the blue, green and red cameras from the top to the bottom, respectively]{Resulting PD of HD~14069 for five time span ranges for the blue, green and red cameras from the top to the bottom, respectively. Solid lines correspond to the mean values, while dotted lines to one standard deviation in each single panel.}\label{fig:hd14_pd}
\end{figure}

\end{knitrout}

\label{lastpage}
\end{document}